\newcommand{\idty}{{\leavevmode{\rm 1\mkern -5.4mu I}}}
\newcommand{\Ir}{Z\!\!\!Z}
\newcommand{\Ibb}[1]{ {\rm I\ifmmode\mkern
            -3.6mu\else\kern -.2em\fi#1}}
\newcommand{\ibb}[1]{\leavevmode\hbox{\kern.3em\vrule
     height 1.2ex depth -.3ex width .2pt\kern-.3em\rm#1}}
\newcommand{\Cx}{{\ibb C}}
\newcommand{\Rl}{{\Ibb R}}
\documentstyle[preprint,aps]{revtex}
\tighten
\begin{document}
\draft
\renewcommand{\theequation} {\arabic{section}.\arabic{equation}}
\preprint{GOET-TP 1/94, revised}
\title{\huge Discrete Differential Calculus, \\
       Graphs, Topologies and Gauge Theory}

\author{Aristophanes Dimakis}
\vskip.3cm
\address{Department of Mathematics, University of Crete,
      GR-71409 Iraklion, Greece}
\author{Folkert M\"uller-Hoissen}
\vskip.3cm
\address{Institut f\"ur Theoretische Physik,
      Bunsenstr. 9, D-37073 G\"ottingen, Germany}

\author{to appear in {\em J. Math. Phys.}}

\maketitle

\begin{abstract}
\noindent
Differential calculus on discrete sets is developed in the spirit of
noncommutative geometry. Any differential algebra on a discrete set can
be regarded as a `reduction' of the `universal differential algebra'
and this allows a systematic exploration of differential algebras
on a given set. Associated with a differential algebra is a
(di)graph where two vertices are connected by at most two
(antiparallel) arrows. The interpretation of such a graph as a
`Hasse diagram' determining a (locally finite) topology
then establishes contact with recent work by other authors in
which discretizations of topological spaces and corresponding
field theories were considered which retain their global topological
structure. It is shown that field theories, and in particular gauge
theories, can be formulated on a discrete set in close analogy with the
continuum case. The framework presented generalizes ordinary lattice
theory which is recovered from an oriented (hypercubic) lattice graph.
It also includes, e.g., the two-point space used by Connes and Lott (and
others) in models of elementary particle physics. The formalism suggests
that the latter be regarded as an approximation of a manifold and thus
opens a way to relate models with an `internal' discrete space ({\`a} la
Connes et al.) to models of dimensionally reduced gauge fields.
Furthermore, also a `symmetric lattice' is studied which (in a certain
continuum limit) turns out to be related to a `noncommutative
differential calculus' on manifolds.
\end{abstract}


\section{Introduction}
\setcounter{equation}{0}
In the context of `noncommutative geometry' \cite{Conn86,Coqu89} a
generalization of the notion of differential forms (on a manifold) plays
a crucial role. With any associative algebra $\cal A$ (over $\Rl$ or
$\Cx$) one associates a {\em differential algebra} which is a
$\Ir$-graded associative algebra $\Omega({\cal A}) = \bigoplus_{r=0}
^\infty \Omega^r({\cal A})$ (where $\Omega^r({\cal A})$ are
$\cal A$-bimodules and $\Omega^0({\cal A})= \cal A$) together with a
linear operator $d \, : \, \Omega^r({\cal A}) \rightarrow \Omega^{r+1}
({\cal A})$ satisfying $ d^2=0$ and $ d(\omega \omega') = (d\omega) \,
\omega'+(-1)^r \omega \, d\omega' $ where $\omega \in \Omega^r
({\cal A})$.
This structure has been studied for
{\em non}-commutative algebras in many recent papers (in particular,
for quantum groups, see \cite{MH+Reut93} and the references given
there). But even {\em commutative} algebras, i.e. algebras of functions
on some topological space, are very much of interest in this respect for
mathematics and physics. A particular example is provided
\cite{Conn+Lott90} by models of elementary particle physics with an
extended space-time of the form $M \times \Ir_2$ where $M$ is an
ordinary four-dimensional space-time manifold. The extension of
differential calculus to discrete spaces allows a corresponding
extension of the Yang-Mills action to $M \times \Ir_2$ which
incorporates Higgs fields and the usual Higgs potential.
Our work in \cite{DMH92,DMHS93}
can be viewed as a lattice analogue of the $\Ir_2$ calculus.
In \cite{DMH93-fs} we went beyond
lattices to an exploration of differential calculus and gauge
theory on arbitrary finite or countable sets. In particular,
some ideas about `reductions' of the {\em universal} differential
algebra (the `universal differential envelope' of $\cal A$
\cite{Conn86,Coqu89}) arose in that work. The present paper presents
a much more complete treatment of the universal differential algebra,
reductions of it, and gauge theory on discrete (i.e., finite or
countable) sets.
\vskip.3cm

Furthermore, the formalism developed in this paper provides a bridge
between noncommutative geometry and various treatments of field theories
on discrete spaces (like lattice gauge theory). We may view a field
theory on a discrete set as an approximation of a continuum theory,
e.g., for the purpose of numerical simulations. More interesting,
however, is in this context the idea that a discrete space-time
could actually be more fundamental than the continuum. This idea has
been discussed and pursued by numerous authors (see
\cite{discrete-spacetime}, in particular).
\vskip.3cm

Discrete spaces have been used in \cite{Sork91} to approximate
general topological spaces and manifolds, taking their global
topological structure into account (see also \cite{Anez94}).
We establish a relation with noncommutative geometry. In particular,
the two-point space in Connes' model can then be regarded as an
approximation of an `internal' manifold as considered in models of
dimensional reduction of gauge theories (see \cite{Kape+Zoup92} for
a review). The appearance of a Higgs field and a Higgs potential in
Connes' model then comes as no surprise since this is a familiar
feature in the latter context. In 1979 Manton derived the bosonic part
of the Weinberg-Salam model from a 6-dimensional Yang-Mills theory on
(4-dimensional) Minkowski space times a 2-dimensional sphere
\cite{Mant79}.
\vskip.3cm

In section II we introduce differential calculus on a discrete set.
Reductions of the `universal differential algebra' are considered in
section III where we discuss the relation with digraphs and Hasse
diagrams (which assign a topology to the discrete set \cite{Sork91}).
Section IV deals with gauge theory, and in particular the case of the
universal differential algebra. Sections V and VI treat, respectively,
the oriented and the `symmetric' lattice as particular examples
of graphs defining a differential algebra. Finally, section VII
summarizes some of the results and contains further remarks.

\section{Differential calculus on a discrete set}
\setcounter{equation}{0}
We consider a countable set $\cal M$ with elements $i,j, \ldots$.
Although we include the case of infinite sets (in particular when it
comes to lattices) in this work, our calculations are then formal rather
than rigorous (since we simply commute operators with infinite sums, for
example). Either one may regard these cases as idealizations of the case
of a large finite set, or one finally has to work with a representation
of the corresponding differential algebra (as in
\cite{Conn+Lott90,Conn93}, see also \cite{DMH93-fs}).
\vskip.3cm

Let ${\cal A}$ be the algebra of $\Cx$-valued functions on $\cal M$.
Multiplication is defined pointwise, i.e.  $(fh)(i) = f(i) \, h(i)$.
There is a distinguished set of functions $e_i$ on $\cal M$ defined by
$e_i(j) = \delta_{ij}$. They satisfy the relations
\begin{eqnarray}
  e_i \, e_j = \delta_{ij} \, e_i   \qquad , \qquad
   \sum_i e_i= \idty
\end{eqnarray}
where $\idty$ denotes the constant function $\idty(i)=1$.
Each $f\in {\cal A}$ can then be written as
\begin{eqnarray}
             f = \sum_i f(i) \, e_i   \; .
\end{eqnarray}
The algebra ${\cal A}$ can be extended to a $\Ir$-graded {\em differential
algebra} $\, \Omega({\cal A})=\bigoplus_{r=0}^\infty \Omega^r({\cal A})$
(where $\Omega^0({\cal A})= \cal A$) via the action of a linear operator
$d \, : \, \Omega^r({\cal A}) \rightarrow \Omega^{r+1}({\cal A})$
satisfying
\begin{eqnarray}
  d\idty=0 \quad , \quad d^2=0 \quad , \quad
  d(\omega_r \, \omega') = (d\omega_r) \, \omega' +
                           (-1)^r \, \omega_r \, d\omega'
\end{eqnarray}
where $\omega_r \in \Omega^r({\cal A})$. The spaces $\Omega^r({\cal A})$
of {\em $r$-forms} are $\cal A$-bimodules.
$\idty$ is taken to be the unit in $\Omega({\cal A})$.
{}From the above properties of the set of functions $e_i$ we obtain
\begin{eqnarray}
        e_i \, de_j = - (de_i) \, e_j + \delta_{ij} \, de_i
\end{eqnarray}
and
\begin{eqnarray}
                   \sum_i de_i = 0
\end{eqnarray}
(assuming that $d$ commutes with the sum) which shows that the $de_i$
are linearly dependent. Let us define
\begin{eqnarray}
           e_{ij} := e_i \,de_j   \quad (i \neq j)
   \quad , \quad    e_{ii} := 0
\end{eqnarray}
(note that $e_i \, de_i \neq 0$) and
\begin{eqnarray}            \label{e...}
 e_{i_1 \ldots i_r} := e_{i_1 i_2} e_{i_2 i_3} \cdots e_{i_{r-1}i_r}
\end{eqnarray}
which for $i_k \neq i_{k+1}$ equals $e_{i_1} \, de_{i_2} \cdots
de_{i_r} $. Then
\begin{eqnarray}       \label{e-e}
 e_{i_1 \ldots i_r} \, e_{j_1 \ldots j_s} = \delta_{i_r j_1} \,
        e_{i_1 \ldots i_{r-1} j_1 \ldots j_s}
        \qquad (r,s \geq 1)   \; .
\end{eqnarray}
A simple calculation shows that
\begin{eqnarray}
         de_i = \sum_j (e_{ji}-e_{ij})    \label{de_i}
\end{eqnarray}
and consequently
\begin{eqnarray}
         df = \sum_{i,j} \, e_{ij} \, (f(j)-f(i))
              \qquad (\forall f\in {\cal A})  \; .      \label{df}
\end{eqnarray}
Furthermore,
\begin{eqnarray}
 de_{ij} & = & de_ide_j=\sum_{k,\ell} (e_{ki} - e_{ik}) (e_{\ell j}
              -e_{j\ell})                           \nonumber \\
         & = & \sum_{k,\ell} (\delta_{i\ell} \, e_{kij} -
               \delta_{k\ell} \, e_{ikj}
                + \delta_{kj} \, e_{ij\ell} )   \nonumber \\
         & = & \sum_k (e_{kij} - e_{ikj} + e_{ijk})  \; .
                                          \label{de_ij}
\end{eqnarray}
Any 1-form $\rho$ can be written as $\rho=\sum e_{ij} \,
\rho_{ij}$ with $\rho_{ij} \in \Cx$ and $\rho_{ii}=0$. One
then finds
\begin{eqnarray}
 d\rho = \sum_{i,j,k} e_{ijk} \, (\rho_{jk}-\rho_{ik}+\rho_{ij}) \; .
\end{eqnarray}
More generally, we have
\begin{eqnarray}
  de_{i_1 \ldots i_r}
  & = &  \sum_j\sum_{k=1}^{r+1} (-1)^{k+1} e_{i_1 \ldots i_{k-1}j i_k
         \ldots i_r}                             \nonumber \\
  & = & \sum_{j_1,\ldots,j_{r+1}} e_{j_1 \ldots j_{r+1}} \,
        \sum_{k=1}^{r+1} (-1)^{k+1} \, I^{j_1 \ldots \widehat{j_k}
        \ldots j_{r+1}}_{i_1 \ldots \ldots \ldots \ldots i_r}
\label{drf}
\end{eqnarray}
where a hat indicates an omission and
\begin{eqnarray}
   I^{j_1 \ldots j_r}_{i_1 \ldots i_r} :=
               \delta^{j_1}_{i_1} \cdots \delta^{j_r}_{i_r}  \; .
\end{eqnarray}
Any $\psi \in \Omega^{r-1}({\cal A})$ can be written as
\begin{eqnarray}
 \psi = \sum_{i_1, \ldots , i_r} e_{i_1 \ldots i_r} \,
        \psi_{i_1 \ldots i_r}
\end{eqnarray}
with $\psi_{i_1\ldots i_r} \in \Cx$, $\psi_{i_1 \ldots i_r} = 0$
if $i_s = i_{s+1}$ for some $s$. We thus have
\begin{eqnarray}
 d\psi = \sum_{i_1,\ldots,i_{r+1}} e_{i_1 \ldots i_{r+1}} \,
 \sum_{k=1}^{r+1} (-1)^{k+1} \, \psi_{i_1 \ldots \widehat{i_k}
 \ldots i_{r+1}}  \; .   \label{dpsi}
\end{eqnarray}
\vskip.3cm
\noindent
If no further relations are imposed on the differential algebra (see
section III, however), we
call it the {\em universal} differential algebra (it is usually
called `universal differential envelope' of $\cal A$
\cite{Conn86,Coqu89}). This particular case will be considered in the
following. The $e_{ij}$ with $i \neq j$ are then a basis of the
space of 1-forms. More generally, it can be shown that
$e_{i_1 \ldots i_r}$ with $i_k \neq i_{k+1}$ for $k=1,\ldots, r-1$
form a basis (over $\Cx$) of the space of $(r-1)$-forms ($r>1$).
As a consequence, $df=0$ implies $f(i)=f(j)$ for all $i,j \in {\cal M}$,
i.e. $f=\mbox{const.}$.
$d\rho=0$ implies the relation $\rho_{jk}=\rho_{ik}-\rho_{ij}$ for
$i \neq j \neq k$. Hence, we have $\rho_{jk}=\rho_{0k}-\rho_{0j}$
with some fixed element $0 \in {\cal M}$.
With $f:=\sum_i\rho_{0i} \, e_i \in {\cal A}$ we find $\rho=df$.
Hence every closed 1-form is exact.
\vskip.3cm
\noindent
{\em Remark.}
The condition $d\rho=0$ can also be written as
      $ \rho_{ik}=\rho_{ij}+\rho_{jk} $
which has the form of the Ritz-Rydberg combination principle
\cite{Ritz08} for the frequences of atomic spectra (see also
\cite{Conn93}). With $\nu:=\sum_{nm}\nu_{nm}\,e_{nm}$
the Ritz-Rydberg principle can be expressed in the simple form $d\nu=0$
which implies $\nu=dH/h$ with the energy $H:=\sum_n E_n \, e_n$ (and
Planck's constant $h$). The equation of motion
\begin{eqnarray}
 {d \over dt} \, \rho = {i \over \hbar} \, \lbrack H , \rho \rbrack
\end{eqnarray}
for a 1-form $\rho$ (with $t$-dependent coefficients) is equivalent
to
\begin{eqnarray}
   {d \over dt} \, \rho_{ij} = {i \over \hbar} \, ( E_i - E_j )
                               \, \rho_{ij}
\end{eqnarray}
which appeared as an early version of the Heisenberg equation
(see \cite{Born+Jord25}).
                                              \hfill  {\large $\Box$}
\vskip.3cm
\noindent
Using (\ref{dpsi}), $d\psi=0$ implies
\begin{eqnarray}
 \sum_{k=1}^{r+1} (-1)^k \, \psi_{i_1 \ldots \widehat{i_k} \ldots
 i_{r+1}} = 0
\end{eqnarray}
which leads to the expression
\begin{eqnarray}
 \psi_{i_1 \ldots i_r} & = & \sum_{k=1}^r (-1)^{k+1} \,
 \psi_{0 i_1 \ldots \widehat{i_k} \ldots i_r}   \nonumber \\
 & = & \psi_{0 i_2 \ldots i_r}
  -\psi_{0 i_1i_3 \ldots i_r}+ \ldots +(-1)^{r+1}\psi_{0 i_1 \ldots
 i_{r-1}}  \; .
\end{eqnarray}
With
\begin{eqnarray}
 \phi := \sum_{i_1,\ldots,i_{r-1}}\,  e_{i_1 \ldots i_{r-1}}
         \, \psi_{0 i_1 \ldots i_{r-1}}
\end{eqnarray}
we find $d\phi=\psi$. Hence, closed forms are always exact so that
the cohomology of $d$ for the universal differential algebra is
trivial.

\vskip.3cm
\noindent
Now we introduce some general notions which are not
restricted to the case of the universal differential algebra.

\vskip.3cm
\noindent
An {\em inner product} on a differential algebra $\Omega({\cal A})$
should have the properties $(\psi_r,\phi_s) = 0$ for $r \neq s$ and
\begin{eqnarray}
 (\psi,c \, \phi) = c \, (\psi,\phi) \quad (\forall c \in \Cx)
 \quad , \quad
 \overline{(\psi,\phi)} = (\phi,\psi)             \label{ip-rules}
\end{eqnarray}
(a bar indicates complex conjugation). Furthermore, we should
require that $(\psi,\phi) = 0 \; \forall \phi \,$ implies
$\psi = 0$.

\vskip.3cm
\noindent
An {\em adjoint} $d^\ast \, : \, \Omega^r({\cal A}) \rightarrow
\Omega^{r-1}({\cal A})$ of
$d$ with respect to an inner product is then defined by
\begin{eqnarray}            \label{d-ast}
   (\psi_{r-1} , d ^\ast\phi_r) := (d\psi_{r-1} , \phi_r)  \; .
\end{eqnarray}
This allows us to construct a {\em Laplace-Beltrami operator} as
follows,
\begin{eqnarray}
        \Delta := - d \, d^\ast - d^\ast \, d  \; .
\end{eqnarray}

\vskip.3cm
\noindent
Let $\Omega({\cal A})^\ast$ denote the $\Cx$-dual of $\Omega({\cal A})$.
The inner product determines a mapping $\psi \in \Omega({\cal A})
\rightarrow \psi^\natural \in \Omega({\cal A})^\ast$ via
\begin{eqnarray}
 \psi^\natural(\omega) := (\psi,\omega) \qquad \qquad
                (\forall \omega \in \Omega({\cal A}))  \; .
\end{eqnarray}
For $c \in \Cx$ we have $(c \, \psi)^\natural = \overline{c} \,
\psi^\natural$. Now we can introduce products $\bullet \, : \,
\Omega^{r}({\cal A})^\ast \times \Omega^{r+p}({\cal A}) \rightarrow
\Omega^{p}({\cal A})$ and $\bullet \, : \, \Omega^{r+p}({\cal A})
\times \Omega^{p}({\cal A})^\ast \rightarrow \Omega^r({\cal A})$ by
\begin{eqnarray}
 (\phi_p , \psi_r^\natural \bullet \omega_{r+p}) :=
       (\psi_r \phi_p , \omega_{r+p})     \quad , \quad
  (\phi_p , \omega_{r+p} \bullet \psi_r^\natural ) :=
       (\phi_p \psi_r , \omega_{r+p})        \label{bullets}
\end{eqnarray}
($\forall \, \phi_p, \, \omega_{r+p}$).
They have the properties
\begin{eqnarray}
 (\psi \phi)^\natural \bullet \omega & = & \phi^\natural
     \bullet (\psi^\natural \bullet \omega)       \label{2bull} \\
  \omega \bullet (\psi \phi)^\natural & = & (\omega \bullet
    \phi^\natural) \bullet \psi^\natural       \; .
\end{eqnarray}
For an inner product such that
\begin{eqnarray}
 (e_{i_1 \ldots i_r} , e_{j_1 \ldots j_r}) := \delta_{i_1 j_1} \,
      g_{i_1 \ldots i_r j_1 \ldots j_r} \, \delta_{i_r j_r}
                                    \label{innerpr}
\end{eqnarray}
with constants $g_{i_1 \ldots i_r j_1 \ldots j_r}$,
the $\bullet$-products satisfy the relations
\begin{eqnarray}
 f^\natural \bullet \omega &=& \overline{f} \, \omega   \\
 \omega \bullet f^\natural &=& \omega \, \overline{f}  \\
 (\psi \, f)^\natural \bullet \omega & = &
    \overline{f} \, (\psi^\natural \bullet\omega)    \label{ph-i} \\
 \omega \bullet (f \, \psi)^\natural & = & (\omega \bullet \psi^\natural)
                                           \, \overline{f}       \\
 \psi^\natural \bullet (f \omega) & = & (\overline{f} \, \psi)^\natural
                                      \bullet \omega    \\
 (f \, \omega) \bullet \psi^\natural & = & f \, (\omega \bullet
       \psi^\natural)                  \\
 \psi^\natural \bullet (\omega f) & = & (\psi^\natural \bullet \omega)
                                         \, f \\
 (\omega f) \bullet \psi^\natural & = & \omega \bullet
                                        (\psi \bar{f})^\natural  \; .
\end{eqnarray}
Whereas our ordinary product of differential forms corresponds to the
cup-product in algebraic topology, the $\bullet$ is related to the
cap-product of a cochain and a chain \cite{Hock+Youn61}.
\vskip.3cm
\noindent
Let us now turn again to the particular case of the universal
differential algebra on $\cal A$. An inner product is then determined
by (\ref{innerpr}) with
\begin{eqnarray}
 g_{i_1 \ldots i_r j_1 \ldots j_r} := \mu_r \, \delta_{i_1 j_1}
 \cdots \delta_{i_r j_r} \, \sigma_{i_1 \ldots i_r}
\end{eqnarray}
where $\mu_r \in \Rl^+$ and
\begin{eqnarray}
 \sigma_{i_1 \ldots i_r} :=  \prod_{s=1}^{r-1}
                            (1- \delta_{i_s i_{s+1}}) \; .
\end{eqnarray}
The factor $\sigma_{i_1 \ldots i_r}$ takes care of the fact that
$e_{i_1 \ldots i_r}$ vanishes if two neighbouring indices coincide.
We then have
\begin{eqnarray}
 (\psi,\phi) = \mu_r \, \sum_{i_1, \ldots, i_r} \bar{\psi}_{i_1
               \ldots i_r} \, \phi_{i_1 \ldots i_r}
\end{eqnarray}
for $r$-forms $\psi, \phi$.
For the adjoint of $d$, we obtain the formulae
\begin{eqnarray}
  d^\ast e_{i_1 \ldots i_r} = {\mu_r \over \mu_{r-1}} \,
  \sum_{k=1}^r(-1)^{k+1} e_{i_1 \ldots \widehat{i_k} \ldots i_r}
\end{eqnarray}
and
\begin{eqnarray}
  d^\ast \psi_r = {\mu_{r+1} \over \mu_r} \,
  \sum_{i_1, \ldots, i_r} e_{i_1 \ldots i_r} \,
  \sum_j \sum_{k=1}^{r+1} (-1)^{k+1} \, \psi_{i_1 \ldots i_{k-1} j
  i_k \ldots i_r}  \; .
\end{eqnarray}
Obviously, $d^\ast f = 0$ for $f\in \cal A$ and $(d^\ast)^2=0$.
If $\cal M$ is a {\em finite} set of $N$ elements, one finds
\begin{eqnarray}
    \Delta f = 2 \, N \, {\mu_2 \over \mu_1} \,
              ({1 \over N} \mbox{Tr} f - f)
\end{eqnarray}
for $f \in {\cal A}$ where $\mbox{Tr} f := \sum_i f(i) $.
\vskip.3cm
\noindent
For the $\bullet$-products we obtain
\begin{eqnarray}
       e_{i_1 \ldots i_r}^\natural \bullet e_{j_1 \ldots j_s}
 & = & {\mu_s \over \mu_{s-r+1}} \, \delta_{i_1 j_1} \cdots
       \delta_{i_r j_r} \;
       e_{j_r \ldots j_s} \; \sigma_{i_1 \ldots i_r}
                                       \label{bullet-e1}   \\
       e_{j_1 \ldots j_s} \bullet e_{i_1 \ldots i_r}^\natural
 & = & {\mu_s \over \mu_{s-r+1}} \, \delta_{i_1 j_{s-r+1}} \cdots
       \delta_{i_r j_s}
       \; e_{j_1 \ldots j_{s-r+1}} \;  \sigma_{i_1 \ldots i_r}
          \; .                             \label{bullet-e2}
\end{eqnarray}

\vskip.3cm
\noindent
{\em Remark.} There is a representation of the universal differential
algebra such that $df = \idty \otimes f - f \otimes \idty$ (cf
\cite{Coqu89}, for example). From this we obtain
$e_{ij} = e_i \otimes e_j$ for $i\neq j$ and
$e_{i_1 \ldots i_r} = \sigma_{i_1 \ldots i_r} \,
e_{i_1} \otimes \cdots \otimes e_{i_r}$.
Hence, $e_{i_1 \ldots i_r}$ can be regarded as an $r$-linear mapping
${\cal M}^r \rightarrow \Cx$,
\begin{eqnarray}
 \langle e_{i_1 \ldots i_r}, (j_1,\ldots,j_r) \rangle &:=&
  \sigma_{i_1 \ldots i_r}
 \, e_{i_1} \otimes \cdots \otimes e_{i_r} \, (j_1,\ldots,j_r)
                        \quad  \nonumber  \\
  &=&  \sigma_{i_1 \ldots i_r} \, \delta_{i_1 j_1}
      \cdots \delta_{i_r j_r}  \; .
\end{eqnarray}
Obviously, tuples $(j_1,\ldots,j_r)$ with $j_s = j_{s+1}$ for some
$s$ lie in the kernel of this mapping.
We can then introduce {\em boundary} and {\em coboundary}
operators,  $\partial$ and $\partial^\ast$ respectively, on ordered
$r$-tuples of elements of $\cal M$ via
\begin{eqnarray}
 \langle e_{i_1 \ldots i_{r-1}},\partial(j_1,\ldots,j_r) \rangle
 & := & \langle d e_{i_1 \ldots i_{r-1}},(j_1,\ldots,j_r)\rangle  \\
 \langle e_{i_1 \ldots i_r},\partial^\ast(j_1,\ldots,j_{r+1})\rangle
 & := & \langle d^\ast e_{i_1 \ldots i_r},(j_1,\ldots,j_{r+1})\rangle
      \; .
\end{eqnarray}
One finds
\begin{eqnarray}
 \partial(i_1,\ldots,i_r) & = & \sum_{k=1}^r(-1)^{k+1}
         (i_1,\ldots,\widehat{i_k},\ldots,i_r) \\
 \partial^\ast(i_1,\ldots,i_r) & = & \sum_i\sum_{k=1}^{r+1}(-1)^{k+1}
         (i_1,\ldots,i_{k-1},i,i_k,\ldots,i_r)
\end{eqnarray}
and that $\partial^\ast$ is the adjoint of $\partial$ with respect to
the inner product defined by
\begin{eqnarray}
 ((i_1,\ldots,i_r),(j_1,\ldots,j_s)) = \delta_{r,s} \, \delta_{i_1,j_1}
 \cdots \delta_{i_r,j_r}  \; .
\end{eqnarray}
This construction makes contact with simplicial homology theory
\cite{Hock+Youn61}. There, however, $r$-tuples are identified (up to a
sign in case of oriented simplexes) if they differ only by a permutation
of their vertices. In the case under consideration, we have general
$r$-tuples, not subject to any condition at all.
                                          \hfill {\large $\Box$}

\vskip.3cm
\noindent
Let $\, \tilde{ } \,$ denote an involutive mapping of $\cal M$
($\tilde{\tilde{k}}=k$ for all $k \in {\cal M}$). It induces
an {\em involution} ${}^\ast$ on $\Omega({\cal A})$ by requiring
$(\psi \phi)^\ast = \phi^\ast \psi^\ast$, $(d\omega_r)^\ast =
(-1)^r d\omega_r^\ast$ and $(f^\ast)(k)=\overline{f(\tilde{k})}$
where a bar denotes complex conjugation.
Using the Leibniz rule, one finds
\begin{eqnarray}            \label{e-ast}
   e_{k \ell}^\ast = -e_{\tilde{\ell} \tilde{k}} \; .
\end{eqnarray}
A natural involution is induced by the identity
$\tilde{k}=k$. Then $e_{k \ell}^\ast = - e_{\ell k}$ and thus
$e_{i_1 \ldots i_r}^\ast = (-1)^{r+1} \, e_{i_r \ldots i_1}$.

\section{Differential algebras and topologies}
\setcounter{equation}{0}
If one is interested, for example, to approximate a differentiable
manifold by a discrete set, the universal differential algebra is too
large to provide us with a corresponding analogue of the algebra of
differential forms on the manifold. We need `smaller' differential
algebras. The fact that the 1-forms $e_{ij}$ ($i \neq j$) induce
via (\ref{e...}) a basis (over $\Cx$) for the spaces of higher
forms together with the relations (\ref{e-e}) offer a simple
way to {\em reduce} the differential algebra.
Setting some of the $e_{ij}$ to zero does not generate any relations
for the remaining (nonvanishing) $e_{k \ell}$ and is consistent with
the rules of differential calculus. It generates relations for forms
of higher grades, however. In particular, it may require that some
of those $e_{i_1 \ldots i_r}$ with $r>2$ have to vanish which do
not contain as a factor any of the $e_{ij}$ which are set to zero
(cf example 1 below).
\vskip.2cm

The reductions of the universal differential algebra obtained in this
way are conveniently represented by graphs as follows.
We regard the elements of $\cal M$ as vertices and associate with
$e_{ij} \neq 0$ an arrow from $i$ to $j$. The universal differential
algebra then corresponds to the graph where all the vertices
are connected pairwise by arrows in both directions. Deleting
some of the arrows leads to a graph which represents a reduction of
the universal differential algebra.
\vskip.2cm

In the following we discuss some simple examples and establish a
relation with topology (some related aspects are discussed in the
appendix). More complicated examples are presented in sections V and VI.

\vskip.3cm
\noindent
{\em Example 1.} Let us consider a set of three elements with the
differential algebra determined by the graph in Fig.1a.

\unitlength2.cm
\begin{picture}(5.,1.8)(-3.8,-0.4)
\thicklines
\put(0.,1.) {\circle*{0.1}}
\put(1.,0.) {\circle*{0.1}}
\put(-1.,0.) {\circle*{0.1}}
\put(-1.25,-0.05) {$0$}
\put(1.2,-0.05) {$1$}
\put(-0.05,1.2) {$2$}
\put(-1.,0.) {\vector(1,0){1.9}}
\put(1.,0.) {\vector(-1,1){0.9}}
\put(0.,1.) {\vector(-1,-1){0.9}}
\end{picture}

\centerline{
\begin{minipage}[t]{9cm}
\centerline{\bf Fig.1a}
\vskip.1cm \noindent
\small
The graph associated with a differential algebra on a set
of three elements.
\end{minipage}    }

\vskip.4cm
\noindent
The nonvanishing basic 1-forms are then $e_{01}, e_{12}, e_{20}$.
{}From these we can only build the basic 2-forms $e_{012}$, $e_{120}$
and $e_{201}$. However, (\ref{de_ij}) yields
\begin{eqnarray}
 0 = d e_{10} = \sum_{k=1}^3 ( e_{k10} - e_{1k0} + e_{10k} )
   = - e_{120}
\end{eqnarray}
and similarly $e_{012} = 0 = e_{201}$. Hence there are no 2-forms (and
thus also no higher forms). This may be interpreted in such a way
that the differential algebra assigns a {\em one}-dimensional
structure to the three-point set. Using (\ref{de_i}), we have
\begin{eqnarray}
  d e_{0} = e_{20} - e_{01} \quad , \quad
  d e_{1} = e_{01} - e_{12} \quad , \quad
  d e_{2} = e_{12} - e_{20}  \; .
\end{eqnarray}
Let us extend the graph in Fig.1a in the following way
(see Fig.1b). We add
new vertices corresponding to the 1-forms on the respective
edges of the diagram. The arrows from the 0-form vertices to the
1-form vertices are then determined by the last equations.
For example, $e_{01}$ appears with a minus sign on the rhs of the
expression for $de_0$. We draw an arrow from the $e_0$ vertex to
the new $e_{01}$ vertex. $e_{20}$ appears with a plus sign and
we draw an arrow from the $e_{20}$ vertex to the $e_0$ vertex.

\unitlength2.cm
\begin{picture}(5.,1.8)(-3.8,-0.4)
\thicklines
\put(-1.,0.) {\circle*{0.1}}
\put(1.,0.) {\circle*{0.1}}
\put(0.,1.) {\circle*{0.1}}
\put(-0.5,0.5) {\circle{0.1}}
\put(0.5,0.5) {\circle{0.1}}
\put(0.,0.) {\circle{0.1}}
\put(-1.25,-0.05) {$0$}
\put(1.2,-0.05) {$1$}
\put(-0.05,1.2) {$2$}
\put(-.85,0.5) {$20$}
\put(0.7,0.5) {$12$}
\put(-0.1,-.3) {$01$}
\put(-1.,0.) {\vector(1,0){0.9}}
\put(0.1,0.) {\vector(1,0){0.8}}
\put(1.,0.) {\vector(-1,1){0.4}}
\put(0.4,0.6) {\vector(-1,1){0.35}}
\put(0.,1.) {\vector(-1,-1){0.4}}
\put(-0.6,0.4) {\vector(-1,-1){0.35}}
\end{picture}

\centerline{
\begin{minipage}[t]{9cm}
\centerline{\bf Fig.1b}
\vskip.1cm \noindent
\small
The extension of the graph in Fig.1a obtained from the latter by
adding new vertices corresponding to the nonvanishing basic 1-forms.
\end{minipage}    }

\vskip.4cm
\noindent
Another form of the graph is shown in Fig.1c where vertices
corresponding to differential forms with the same grade are grouped
together horizontally and $(r+1)$-forms are below $r$-forms.

\unitlength2.cm
\begin{picture}(5.,1.8)(-3.8,-0.4)
% Hasse diagram, Ex.1
\thicklines
\put(-1.,1.) {\circle*{0.1}}
\put(0.,1.) {\circle*{0.1}}
\put(1.,1.) {\circle*{0.1}}
\put(-1.04,1.2) {$0$}
\put(-0.04,1.2) {$1$}
\put(0.96,1.2) {$2$}
\put(-1.,0.) {\circle{0.1}}
\put(0.,0.) {\circle{0.1}}
\put(1.,0.) {\circle{0.1}}
\put(-1.1,-0.3) {$01$}
\put(-0.1,-0.3) {$12$}
\put(0.9,-0.3) {$20$}
\put(-1.,1.) {\vector(0,-1){0.9}}
\put(0.,1.) {\vector(0,-1){0.9}}
\put(1.,1.) {\vector(0,-1){0.9}}
\put(-1.,0.) {\vector(1,1){0.9}}
\put(0.,0.) {\vector(1,1){0.9}}
\put(1.,0.) {\vector(-2,1){1.9}}
\end{picture}

\centerline{
\begin{minipage}[t]{9cm}
\centerline{\bf Fig.1c}
\vskip.1cm \noindent
\small
The (oriented) Hasse diagram derived from the graph in Fig.1a.
\end{minipage}    }

\vskip.4cm
\noindent
The result can be interpreted
as a {\em Hasse} diagram which determines a finite topology in the
following way \cite{Sork91}. A vertex together with all lower lying
vertices which are connected to it forms an open set. In the present
case, $\lbrace 01 \rbrace, \lbrace 12 \rbrace, \lbrace 20 \rbrace,
\lbrace 0,01,20 \rbrace, \lbrace 1, 01, 12 \rbrace,
\lbrace 2, 12, 20 \rbrace$  are the open sets (besides the empty
and the whole set). This is an approximation to the topology of $S^1$.
It consists of a chain of three open sets covering $S^1$ which already
displays the global topology of $S^1$. In particular, the fundamental
group $\pi_1$ is the same as for $S^1$.

\vskip.4cm
\noindent
In the above example we have defined the {\em dimension} of a
differential algebra as the grade of its highest nonvanishing forms.
This is probably the most fruitful such concept (others have been
considered in \cite{DMH93-fs,Evak94}). Applying it to subgraphs leads
to a {\em local} notion of dimension.

\vskip.4cm
\noindent
{\em Example 2.} Again, we consider a set of three elements, but this
time with the differential algebra determined by the graph in Fig.2a.
In this case, the nonvanishing basic 1-forms are $e_{01}, e_{12},
e_{02}$. From these one can only build the 2-form $e_{012} \,$.
(\ref{de_ij}) then reads
\begin{eqnarray}        \label{de_ij-Ex2}
        d e_{ij} =  e_{0ij} - e_{i1j} + e_{ij2}
\end{eqnarray}
and $e_{012}$ remains as a non-vanishing 2-form.

\unitlength2.cm
\begin{picture}(5.,1.8)(-3.8,-0.4)
\thicklines
\put(0.,1.) {\circle*{0.1}}
\put(1.,0.) {\circle*{0.1}}
\put(-1.,0.) {\circle*{0.1}}
\put(-1.25,-0.05) {$0$}
\put(1.2,-0.05) {$1$}
\put(-0.05,1.2) {$2$}
\put(-1.,0.) {\vector(1,0){1.9}}
\put(1.,0.) {\vector(-1,1){0.9}}
\put(-1.,0.) {\vector(1,1){0.9}}
\end{picture}

\centerline{
\begin{minipage}[t]{9cm}
\centerline{\bf Fig.2a}
\vskip.1cm \noindent
\small
The graph associated with another differential algebra on the set
of three elements.
\end{minipage}    }

\vskip.4cm
\noindent
There are no
higher forms so that the differential algebra assigns {\em two}
dimensions to the three-point set. Using (\ref{de_i}), we have
\begin{eqnarray}
  d e_{0} = -e_{01} - e_{02} \quad , \quad
  d e_{1} = e_{01} - e_{12}  \quad , \quad
  d e_{2} =  e_{02} + e_{12}  \; .
\end{eqnarray}
The extended graph is shown in Fig.2b. Now we have an additional
vertex corresponding to the two-form $e_{012}$ with connecting
arrows determined by (\ref{de_ij-Ex2}).

\unitlength2.cm
\begin{picture}(5.,1.8)(-3.8,-0.4)
\thicklines
\put(-1.,0.) {\circle*{0.1}}
\put(1.,0.) {\circle*{0.1}}
\put(0.,1.) {\circle*{0.1}}
\put(-0.5,0.5) {\circle{0.1}}
\put(0.5,0.5) {\circle{0.1}}
\put(0.,0.) {\circle{0.1}}
\put(0.,0.4) {\circle{0.1}}
\put(0.,0.4) {\circle*{0.05}}
\put(-1.25,-0.05) {$0$}
\put(1.2,-0.05) {$1$}
\put(-0.05,1.2) {$2$}
\put(-.85,0.5) {$02$}
\put(0.7,0.5) {$12$}
\put(-0.1,-.3) {$01$}
\put(-0.13,.5) {\small $012$}
\put(-1.,0.) {\vector(1,0){0.9}}
\put(0.1,0.) {\vector(1,0){0.8}}
\put(1.,0.) {\vector(-1,1){0.4}}
\put(0.4,0.6) {\vector(-1,1){0.35}}
\put(-1.,0.) {\vector(1,1){0.4}}
\put(-0.4,0.6) {\vector(1,1){0.35}}
\put(0.1,0.42) {\vector(4,1){0.34}}
\put(-0.5,0.5) {\vector(4,-1){0.4}}
\put(0.,0.3) {\vector(0,-1){0.2}}
\end{picture}

\centerline{
\begin{minipage}[t]{9cm}
\centerline{\bf Fig.2b}
\vskip.1cm \noindent
\small
The extension of the graph in Fig.2a with new vertices corresponding
to nonvanishing forms of grade higher than zero.
\end{minipage}    }

\vskip.4cm
\noindent
The Hasse diagram is drawn in Fig.2c. The open sets of the
cor\-re\-spon\-ding to\-po\-lo\-gy are
$\lbrace 012 \rbrace, \lbrace 01, 012 \rbrace, \lbrace 12, 012 \rbrace,
\lbrace 02, 012 \rbrace, \lbrace 0, 01, 02, 012 \rbrace,
\lbrace 1, 01, 12, 012 \rbrace, \lbrace 2, 12, 02, 012 \rbrace$.

%\vskip.6cm

\unitlength2.cm
\begin{picture}(5.,2.8)(-3.8,-1.4)
% Hasse diagram, Ex.2
\thicklines
\put(-1.,1.) {\circle*{0.1}}
\put(0.,1.) {\circle*{0.1}}
\put(1.,1.) {\circle*{0.1}}
\put(-1.04,1.2) {$0$}
\put(-0.04,1.2) {$1$}
\put(0.96,1.2) {$2$}
\put(-1.,0.) {\circle{0.1}}
\put(0.,0.) {\circle{0.1}}
\put(1.,0.) {\circle{0.1}}
\put(-1.1,-0.3) {$01$}
\put(-0.3,-0.3) {$12$}
\put(0.9,-0.3) {$02$}
\put(0.,-1.) {\circle{0.1}}
\put(0.,-1.) {\circle*{0.05}}
\put(-0.1,-1.3) {$012$}
\put(-1.,1.) {\vector(0,-1){0.9}}
\put(0.,1.) {\vector(0,-1){0.85}}
\put(1.,0.) {\vector(0,1){0.9}}
\put(-1.,0.) {\vector(1,1){0.9}}
\put(0.1,0.1) {\vector(1,1){0.8}}
\put(-1.,1.) {\vector(2,-1){1.9}}
\put(0.,-0.9) {\vector(0,1){0.8}}
\put(-0.1,-0.9) {\vector(-1,1){0.8}}
\put(1.,0.) {\vector(-1,-1){0.9}}
\end{picture}

\centerline{
\begin{minipage}[t]{9cm}
\centerline{\bf Fig.2c}
\vskip.1cm \noindent
\small
The (oriented) Hasse diagram derived from the graph in Fig.2a.
\end{minipage}    }

\vskip.4cm

%\vskip.6cm

\noindent
The topology is shown in Fig.2d.

%\vskip.6cm

\unitlength2.cm
\begin{picture}(5.,1.8)(-3.8,-0.6)
% topology, Ex.2
\thicklines
\put(-0.2,0.) {\circle{2.}}
\put(0.,0.35) {\circle{2.}}
\put(0.2,0.) {\circle{2.}}
\end{picture}

\centerline{
\begin{minipage}[t]{9cm}
\centerline{\bf Fig.2d}
\vskip.1cm \noindent
\small
The topology on the three point set determined by the Hasse
diagram in Fig.2c.
\end{minipage}    }

\vskip.5cm
\noindent
{\em Example 3.} We supply a set of four elements with the
differential algebra determined by the graph in Fig.3a.

\unitlength2.cm
\begin{picture}(5.,2.7)(-3.8,-1.3)
\thicklines
\put(0.,1.) {\circle*{0.1}}
\put(1.,0.) {\circle*{0.1}}
\put(-1.,0.) {\circle*{0.1}}
\put(0.,-1.) {\circle*{0.1}}
\put(-1.25,-0.05) {$3$}
\put(-.05,-1.3) {$0$}
\put(1.2,-0.05) {$1$}
\put(-0.05,1.2) {$2$}
\put(1.,0.) {\vector(-1,0){1.9}}
\put(1.,0.) {\vector(-1,1){0.9}}
\put(0.,-1.) {\vector(-1,1){0.9}}
\put(0.,-1.) {\vector(1,1){0.9}}
\put(0.,1.) {\vector(-1,-1){0.9}}
\put(0.,-1.) {\vector(0,1){1.9}}
\end{picture}

\centerline{
\begin{minipage}[t]{9cm}
\centerline{\bf Fig.3a}
\vskip.1cm \noindent
\small
A graph which determines a differential algebra on a set of
four points.
\end{minipage}    }

\vskip.5cm
\noindent
The nonvanishing basic 1-forms are thus $e_{01}, e_{02}, e_{03},
e_{12}, e_{13}, e_{23}$. From these we can only build the 2-forms
$e_{012}, e_{013}, e_{023}, e_{123}$ and the 3-form $e_{0123}$.
There are no higher forms. (\ref{drf}) yields
\begin{eqnarray}
  d e_{01} =  e_{012} + e_{013}  \quad , \quad
  d e_{02} = - e_{012} + e_{023}   \quad , \quad
  d e_{12} =  e_{012} + e_{123}  \nonumber \\
  d e_{13} =  e_{013} - e_{123}  \quad , \quad
  d e_{03} = - e_{013} - e_{023}   \quad , \quad
  d e_{23} =  e_{123} + e_{023}
\end{eqnarray}
and
\begin{eqnarray}
  d e_{012} =  - e_{0123}      \quad , \quad
  d e_{013} =   e_{0123}    \quad , \quad
  d e_{023} =  - e_{0123}      \quad , \quad
  d e_{123} =   e_{0123}     \; .
\end{eqnarray}
The corresponding (oriented) Hasse diagram is drawn in Fig.3b.
It determines a topology which approximates the topology of a simply
connected open set in $\Rl^3$.

\unitlength2.cm
\begin{picture}(5.,4.2)(-1.3,-2.5)
% Hasse diagram, Ex.3
\thicklines
\put(1.,1.) {\circle*{0.1}}
\put(2.,1.) {\circle*{0.1}}
\put(3.,1.) {\circle*{0.1}}
\put(4.,1.) {\circle*{0.1}}
\put(0.96,1.2) {$0$}
\put(1.96,1.2) {$1$}
\put(2.96,1.2) {$2$}
\put(3.96,1.2) {$3$}
\put(0.,0.) {\circle{0.1}}
\put(1.,0.) {\circle{0.1}}
\put(2.,0.) {\circle{0.1}}
\put(3.,0.) {\circle{0.1}}
\put(4.,0.) {\circle{0.1}}
\put(5.,0.) {\circle{0.1}}
\put(-0.3,-0.05) {$01$}
\put(0.7,-0.05) {$02$}
\put(1.7,-0.05) {$03$}
\put(3.1,-0.05) {$12$}
\put(4.1,-0.05) {$13$}
\put(5.1,-0.05) {$23$}
\put(1.,-1.) {\circle*{0.1}}
\put(2.,-1.) {\circle*{0.1}}
\put(3.,-1.) {\circle*{0.1}}
\put(4.,-1.) {\circle*{0.1}}
\put(0.7,-1.2) {$012$}
\put(1.7,-1.2) {$013$}
\put(3.1,-1.2) {$023$}
\put(4.1,-1.2) {$123$}
\put(2.5,-2.) {\circle{0.1}}
\put(2.3,-2.3) {$0123$}
\put(1.,1.) {\vector(-1,-1){0.9}}
\put(1.,1.) {\vector(0,-1){0.9}}
\put(1.,1.) {\vector(1,-1){0.9}}
\put(2.,1.) {\vector(1,-1){0.9}}
\put(2.,1.) {\vector(2,-1){1.9}}
\put(3.,1.) {\vector(2,-1){1.9}}
\put(0.,0.) {\vector(2,1){1.9}}
\put(1.,0.) {\vector(2,1){1.9}}
\put(1.,0.) {\vector(0,-1){0.9}}
\put(2.,0.) {\vector(2,1){1.9}}
\put(2.,0.) {\vector(0,-1){0.9}}
\put(2.,0.) {\vector(1,-1){0.9}}
\put(3.,0.) {\vector(0,1){0.9}}
\put(4.,0.) {\vector(0,1){0.9}}
\put(4.,0.) {\vector(0,-1){0.9}}
\put(5.,0.) {\vector(-1,1){0.9}}
\put(1.,-1.) {\vector(-1,1){0.9}}
\put(1.,-1.) {\vector(2,1){1.9}}
\put(1.,-1.) {\vector(3,-2){1.4}}
\put(2.,-1.) {\vector(-2,1){1.9}}
\put(2.,-1.) {\vector(2,1){1.9}}
\put(3.,-1.) {\vector(-2,1){1.9}}
\put(3.,-1.) {\vector(2,1){1.9}}
\put(3.,-1.) {\vector(-1,-2){0.45}}
\put(4.,-1.) {\vector(-1,1){0.9}}
\put(4.,-1.) {\vector(1,1){0.9}}
\put(2.5,-2.) {\vector(3,2){1.4}}
\put(2.5,-2.) {\vector(-1,2){0.45}}
\end{picture}

\centerline{
\begin{minipage}[t]{9cm}
\centerline{\bf Fig.3b}
\vskip.1cm \noindent
\small
The (oriented) Hasse diagram derived from the graph in Fig.3a.
\end{minipage}    }

\vskip.4cm
\noindent
More generally, we have the possibility of reductions on the level of
$r$-forms ($r \geq 1$), i.e. we can set any of the (not already
vanishing) $e_{i_1 \ldots i_{r+1}}$ to zero. In example 2, we have the
freedom to set the 2-form $e_{012}$ to zero by hand. We then end up with
the topology of a circle (as in example 1) instead of the topology of a
disc. Such a `higher-order reduction' has no effect on the remaining
$s$-forms with $s \leq r$, but influences forms of grade higher than
$r$. The full information about the differential algebra is then only
contained in the extended graph in which {\em all} the nonvanishing
basic forms -- and not just the 0-forms -- are represented as vertices.
This is the (oriented) Hasse diagram.
\vskip.3cm
\noindent
{\em Example 4.} The graph in Fig.4a corresponds to the universal
differential algebra on a set of two elements. We
are allowed to omit the arrows since both directions are present.
{}From the basic 1-forms $e_{01}$ and $e_{10}$ we can construct
forms $e_{010101 \ldots}$ and $e_{101010 \ldots}$ of arbitrary
grade.

\unitlength2.cm
\begin{picture}(5.,1.6)(-2.3,-0.9)
\thicklines
\put(1.,0.) {\circle*{0.1}}
\put(2.,0.) {\circle*{0.1}}
\put(1.5,0.) {\oval(1.,.7)[t]}
\put(1.5,0.) {\oval(1.,.7)[b]}
\put(0.8,-0.1) {$0$}
\put(2.1,-0.1) {$1$}
\end{picture}

\centerline{
\begin{minipage}[t]{9cm}
\centerline{\bf Fig.4a}
\vskip.1cm \noindent
\small
The graph associated with the universal differential algebra on
a two point set.
\end{minipage}    }

\vskip.4cm
\noindent
The associated (oriented) Hasse diagram is shown in Fig.4b. If we
set the 2-forms $e_{010}$ and $e_{101}$ to zero, the Hasse diagram
determines a topology which approximates the topology of the
circle $S^1$. If, however, we set the 3-forms $e_{0101}$ and
$e_{1010}$ to zero, an approximation of $S^2$ is obtained, etc..
In this way contact is made with the work in \cite{BBET93}.

\unitlength2.cm
\begin{picture}(5.,3.4)(-3.3,-2.)
% Hasse diagram, Ex.4
\thicklines
\put(0.,1.) {\circle*{0.1}}
\put(1.,1.) {\circle*{0.1}}
\put(-.04,1.2) {$0$}
\put(0.95,1.2) {$1$}
\put(0.,0.) {\circle{0.1}}
\put(1.,0.) {\circle{0.1}}
\put(-0.3,-0.1) {$01$}
\put(1.1,-0.1) {$10$}
\put(0.,-1.) {\circle*{0.1}}
\put(1.,-1.) {\circle*{0.1}}
\put(-0.4,-1.1) {$010$}
\put(1.1,-1.1) {$101$}
\put(0.,1.) {\vector(0,-1){0.9}}
\put(1.,1.) {\vector(0,-1){0.9}}
\put(0.1,0.1) {\vector(1,1){0.8}}
\put(0.9,0.1) {\vector(-1,1){0.8}}
\put(0.,-1.) {\vector(0,1){0.9}}
\put(1.,-1.) {\vector(0,1){0.9}}
\put(0.,-1.) {\vector(1,1){0.9}}
\put(1.,-1.) {\vector(-1,1){0.9}}
\put(0.,-1.2) {\circle*{0.03}}
\put(1.,-1.2) {\circle*{0.03}}
\put(0.,-1.4) {\circle*{0.03}}
\put(1.,-1.4) {\circle*{0.03}}
\put(0.,-1.6) {\circle*{0.03}}
\put(1.,-1.6) {\circle*{0.03}}
\end{picture}

\centerline{
\begin{minipage}[t]{9cm}
\centerline{\bf Fig.4b}
\vskip.1cm \noindent
\small
The (oriented) Hasse diagram for the universal differential
algebra on a set of two elements.
\end{minipage}    }
\vskip.4cm
\noindent
The two-point space can thus be regarded, in particular, as an
approximation of the two-dimensional sphere. As an `internal
space' the latter appears, for example, in Manton's six-dimensional
Yang-Mills model from which he obtained the bosonic sector of the
Weinberg-Salam model by dimensional reduction \cite{Mant79}. In view
of this relation, the appearance of the Higgs field in recent models
of noncommutative geometry {\`a} la Connes and Lott \cite{Conn+Lott90}
(see also section IV) may be traced back essentially to the
abovementioned old result.

Let $e:= e_0$. Then $e_1 = \idty - e$, $e^2 = e$ and $(de) \, e + e \,
de = de$. Comparison with Appendix B in \cite{Kast88} (see also
\cite{Conn86}) shows that we are dealing with the differential envelope
of the complex numbers.
\vskip.4cm
\noindent
{\em Example 5.} Let us consider the `symmetric' graph in Fig.5. The
nonvanishing basic 1-forms are $e_{01}, e_{10}, e_{12}, e_{21}$.
{}From these one obtains the 2-forms $e_{010}, e_{012}, e_{101},
e_{210}, e_{121}, e_{212}$. As a consequence of $e_{02} = 0 = e_{20}$
we find $e_{012} = 0 = e_{210}$. The only nonvanishing basic 2-forms
are thus $e_{010}, e_{101}$ and $e_{121}, e_{212}$.

\unitlength2.cm
\begin{picture}(5.,1.8)(-1.7,-1.)
\thicklines
\put(1.,0.) {\circle*{0.1}}
\put(2.,0.) {\circle*{0.1}}
\put(3.,0.) {\circle*{0.1}}
\put(1.5,0.) {\oval(1.,.7)[t]}
\put(1.5,0.) {\oval(1.,.7)[b]}
\put(2.5,0.) {\oval(1.,.7)[t]}
\put(2.5,0.) {\oval(1.,.7)[b]}
\put(0.8,-0.1) {$0$}
\put(1.8,-0.1) {$1$}
\put(3.1,-0.1) {$2$}
\end{picture}

\centerline{
\begin{minipage}[t]{9cm}
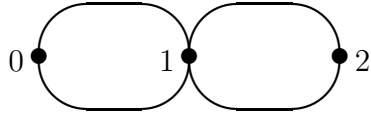
\centerline{\bf Fig.5}
\vskip.1cm \noindent
\small
A `symmetric' differential algebra on a three point set.
\end{minipage}    }

\vskip.4cm
\noindent
In terms of the (coordinate) function $x := 0 \, e_0 + 1 \, e_1
+ 2 \, e_2 = e_1 + 2 \, e_2$ we obtain
\begin{eqnarray}
   e_0 = {\bf 1} - {3 \over 2} \, x + {1 \over 2} \, x^2  \quad ,
   \quad  e_1 = 2 \, x - x^2 \quad , \quad e_2 = {1 \over 2} \,
   (x^2 -x)
\end{eqnarray}
and
\begin{eqnarray}
   \lbrack x , dx \rbrack = - ( e_{01} + e_{10} + e_{12} + e_{21} )
         =: \tau    \; .
\end{eqnarray}
The 1-forms $dx$ and $\tau$ constitute a basis of $\Omega^1({\cal A})$
as a left (or right) $\cal A$-module.
A simple calculation shows that a 1-form $\omega$ is closed iff it
can be written in the form
\begin{eqnarray}
       \omega = c_1 \, (e_{01} - e_{10}) + c_2 \,
                (e_{12} - e_{21})
\end{eqnarray}
with complex constants $c_1, c_2$. Furthermore, closed 1-forms are
exact. The 1-form $\tau$ introduced above is not closed.
\vskip.4cm

A differential algebra with the property that $e_{ij} = 0$ for some
$i,j \in {\cal M}$ only if also $e_{ji} = 0$ is called a {\em symmetric
reduction} of the universal differential algebra. The associated
graph will then also be called {\em symmetric}. The algebra
considered in example 5 is of this type.
\vskip.3cm

In the examples treated above, we started from a differential
algebra and ended up with a topology. One can go the other way
round, i.e. start with a topology, construct the corresponding
Hasse diagram and add directions to its edges in accordance with
the rules of differential calculus (cf \cite{BBET93}).

\section{Gauge theory on a discrete set}
\setcounter{equation}{0}
A field $\Psi$ on $\cal M$ is a cross section of a vector bundle over
$\cal M$, e.g., a cross section of the trivial bundle
${\cal M} \times \Cx^n$. In the
algebraic language the latter corresponds to the (free) $\cal A$-module
${\cal A}^n$ (nontrivial bundles correspond to `finite
projective modules'). We regard it as a {\em left} $\cal A$-module
and consider an action $\Psi \mapsto G \, \Psi$
of a (local) gauge group, a subgroup of $GL(n,{\cal A})$ with elements
$G=\sum_i G(i) \, e_i$, on ${\cal A}^n$. This induces on the dual
(right $\cal A$-) module an action $\alpha \mapsto \alpha \, G^{-1}$.

\vskip.3cm
\noindent
Let us introduce {\em covariant exterior derivatives}
\begin{eqnarray}             \label{DPsi}
      D\Psi = d \Psi + A \, \Psi    \quad , \quad
      D\alpha = d \alpha - \alpha \, A
\end{eqnarray}
where $A$ is a 1-form. These expressions are indeed covariant if
$A$ obeys the usual transformation law of a {\em connection
1-form},
\begin{eqnarray}                            \label{A-transf}
          A' = G \, A \, G^{-1} - dG \, G^{-1}  \; .
\end{eqnarray}
Since $dG$ is a discrete derivative, $A$ cannot be Lie algebra valued.
It is rather an element of $\Omega^1({\cal A}) \otimes_{\cal A}
M_n({\cal A})$ where $M_n({\cal A})$ is the space of
$n \times n$ matrices with entries in $\cal A$.
\vskip.3cm

As a consequence of (\ref{DPsi}) we have
$d(\alpha \, \Psi) = (D \alpha) \, \Psi + \alpha \, (D\Psi)$.
We could have used different connections for the module ${\cal A}^n$
and its dual. The requirement of the last relation would then identify
both.
\vskip.3cm

We call an element $U \in \Omega^1({\cal A}) \otimes_{\cal A}
GL(n, {\cal A})$ a {\em transport operator} (the reason will become
clear in the following) if it transforms as $U \mapsto G \, U \,
G^{-1}$ under a gauge transformation. Since $U = \sum_{i,j} e_{ij} \,
U_{ij}$ with $U_{ij} \in GL(n,\Cx)$, we find
\begin{eqnarray}
       U'_{ij} = G(i) \, U_{ij} \, G(j)^{-1}  \; .
\end{eqnarray}
Using (\ref{A-transf}), (\ref{e-e}) and (\ref{df}) it can be shown that
such a transport operator is given by
\begin{eqnarray}
      U := \sum_{i,j}  e_{ij} \, ( {\bf 1} + A_{ij} )
\end{eqnarray}
where ${\bf 1}$ is the identity in the group.

\vskip.3cm
\noindent
The covariant derivatives introduced above can now be written as
follows,
\begin{eqnarray}
 D\Psi = \sum_{i,j} e_{ij} \, \nabla_j \Psi(i)
 \quad , \quad
 \nabla_j \Psi(i) := U_{ij} \, \Psi(j) - \Psi(i)    \\
 D\alpha = \sum_{i,j} e_{ij}\, \nabla_j \alpha(i)\, U_{ij}
                    \quad , \quad
 \nabla_j \alpha(i) := \alpha(j) \, U_{ij}^{-1} - \alpha(i)
\end{eqnarray}
where $\Psi = \sum_i e_i \, \Psi(i)$ and in the last equation we have
made the additional assumption that $U_{ij}$ is invertible.

\vskip.3cm
\noindent
The 1-forms $e_{ij}$ are linearly independent, except for those which
are set to zero in a reduction of the universal differential algebra.
The condition of covariantly constant $\Psi$, i.e. $D \Psi = 0$, thus
implies $\nabla_j \Psi(i) = 0$ for those
$i,j$ for which $e_{ij} \neq 0$. This gives $U_{ij}$ the interpretation
of an operator for parallel transport from $j$ to $i$. Drawing points
for the elements of $\cal M$, we may assign to $U_{ij}$ an arrow from
$j$ to $i$ (see Fig.6).

\unitlength2.cm
\begin{picture}(5.,1.8)(-3.8,-0.4)
\thinlines
\put(0.,1.) {\circle*{0.1}}
\put(1.,0.) {\circle*{0.1}}
\put(-1.,0.) {\circle*{0.1}}
\put(-1.,0.) {\vector(1,0){1.9}}
\put(1.,0.) {\vector(-1,1){0.9}}
\put(0.,1.) {\vector(-1,-1){0.9}}
\put(-1.2,0.) {$i$}
\put(-0.05,1.2) {$k$}
\put(1.2,0.) {$j$}
\put(-.1,-.3) {$U_{ji}$}
\put(.6,.5) {$U_{kj}$}
\put(-.85,.5) {$U_{ik}$}
\end{picture}

\centerline{
\begin{minipage}[t]{9cm}
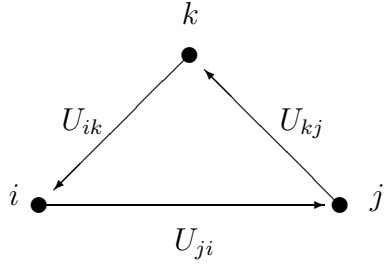
\centerline{\bf Fig.6}
\vskip.1cm \noindent
\small
A visualization of $U_{ij}$ as a transport operator from $j$ to $i$.
\end{minipage}    }

\vskip.4cm
\noindent
The curvature of the connection $A$ is given by the familiar formula
\begin{eqnarray}
             F = dA + A^2
\end{eqnarray}
and transforms in the usual way, $F \mapsto G \, F \, G^{-1}$. As a
2-form, it can be written as $F=\sum_{i,j,k} e_{ijk} \, F_{ijk}$ and
we find
\begin{eqnarray}
     F = \sum_{i,j,k} e_{ijk} (U_{ij} \, U_{jk} - U_{ik})   \; .
                                  \label{F-U}
\end{eqnarray}
\vskip.3cm
\noindent
{\em Remark.} In general, the 2-forms $e_{ijk}$ are not linearly
independent for a given differential algebra so that $F = 0$ does not
imply the vanishing of the coefficients. The latter is true, however,
for the universal differential algebra. In that case vanishing $F$
leads to $U_{ij} \, U_{jk} = U_{ik}$ and, in particular,
$U_{ji} = U_{ij}^{-1}$. As a consequence, $U$ is then
`path-independent' on $\cal M$. If we set $G(i) := U_{i0}$, the
condition of vanishing curvature implies $U_{ij} = G(i) \,
G(j)^{-1}$ which can also be expressed as $A = - dG \, G^{-1}$,
i.e., the connection is `pure gauge'. The Bianchi identity
$DF = dF + [A,F] = 0$ is a 3-form relation. But only for the
universal differential algebra we can conclude that the coefficients
of the basic 3-forms $e_{ijk \ell}$ vanish which then leads to
$F_{ij \ell} - F_{ik \ell} = F_{ijk} \, U_{k\ell} - U_{ij} \,
 F_{jk \ell} $.                       \hfill   {\Large $\Box$}
\vskip.3cm

In order to generalize an inner product (with the properties
specified in section II) to matrix valued forms, we require
that
\begin{eqnarray}
   (\phi , \psi) = \sum \phi_{i_1 \cdots i_r}^\dagger \,
   (e_{i_1 \cdots i_r}, e_{j_1 \cdots j_s}) \; \psi_{j_1\cdots j_s}
                                           \; .
\end{eqnarray}
Here $\phi_{i_1 \ldots i_r}$ is a matrix with entries in $\Cx$ and
$\phi^\dagger_{i_1 \ldots i_r}$ denotes the hermitian conjugate matrix.
\vskip.3cm
\noindent
The {\em Yang-Mills action}
\begin{eqnarray}
                S_{YM} := \mbox{tr} \, (F,F)
\end{eqnarray}
is then gauge-invariant if $G^\dagger = G^{-1}$.
With these tools we can now formulate gauge
theory, in particular, on the differential algebras (respectively graphs)
considered in the previous section. We will not elaborate these examples
here but only discuss the case of the universal differential algebra in
some detail. Other examples will then be treated in sections V and VI.
\vskip.3cm

The hermitian conjugation of complex matrices can be extended to
matrix valued differential forms via
\begin{eqnarray}
  \phi^\dagger = \sum_{i_1 \ldots i_r} (\phi_{i_1 \ldots i_r}
                \, e_{i_1 \ldots i_r})^\dagger
              = \sum_{i_1 \ldots i_r} \phi^\dagger_{i_1 \ldots i_r}
                \, e_{i_1 \ldots i_r}^\ast
\end{eqnarray}
if an involution is given on $\Omega ({\cal A})$.
A {\em conjugation} ${}^\dagger$ acting on a field $\Psi$ is a map from
a left $\cal A$-module, ${\cal A}^n$ in our case, to the dual right
$\cal A$-module such that $ (\phi \, \Psi)^\dagger = \Psi^\dagger \,
\phi^\dagger$ where $\phi$ is an $(n \times n)$-matrix valued
differential form. Since $G^\dagger = G^{-1}$ this is in accordance with
the transformation rule for $\Psi^\dagger$, i.e. $\Psi^\dagger
\mapsto \Psi^\dagger G^{-1}$.

\subsection{Gauge theory with the universal differential algebra}
Using the inner product introduced in section II on the universal
differential algebra, the {\em Yang-Mills action} becomes
\begin{eqnarray}
  S_{YM} =  \sum_{i,j,k} \mbox{tr} \, (F_{ijk}^\dagger \, F_{ijk})
\end{eqnarray}
(where we have set $\mu_r =1$). Using (\ref{F-U}), we get
\begin{eqnarray}
 S_{YM} = \mbox{tr} \sum_{i,j,k} \, ( U_{jk}^\dagger \, U_{ij}^\dagger
          \, U_{ij} \, U_{jk} - U_{jk}^\dagger \, U_{ij}^\dagger \,
          U_{ik} - U_{ik}^\dagger \, U_{ij} \, U_{jk}
          + U_{ik}^\dagger \, U_{ik} )  \; .
                                               \label{YM-action}
\end{eqnarray}
Variation of the Yang-Mills action with respect to the connection
$A$, making use of (\ref{d-ast}) and (\ref{bullets}), leads to
\begin{eqnarray}
(\delta F, F) = (d \delta A + \delta A \, A + A \, \delta A , F) =
              (\delta A, d^\ast F + A^\natural \bullet F
              + F \bullet A^\natural)
\end{eqnarray}
from which we read off the Yang-Mills equation
\begin{eqnarray}
   d^\ast F + A^\natural \bullet F + F \bullet A^\natural = 0 \; .
                      \label{YM-equation}
\end{eqnarray}
\vskip.3cm
\noindent
In the following we will evaluate some of the formulae given above
with the choice of the natural involution (cf section II) and with
certain additional conditions imposed on the gauge field.
The usual compatibility condition for parallel transport and
conjugation is
\begin{eqnarray}
         (D\Psi)^\dagger = D(\Psi^\dagger)
\end{eqnarray}
which is equivalent to
\begin{eqnarray}
                A^\dagger = - A           \label{cond1}
\end{eqnarray}
and implies $F^\dagger = F$. Using (\ref{e-ast}), (\ref{cond1})
becomes
\begin{eqnarray}
        U_{ij}^\dagger = U_{ji}      \label{U^dagger}
\end{eqnarray}
which implies $F_{ijk}^\dagger = F_{kji}$.
Evaluating the Yang-Mills action (\ref{YM-action}) and the Yang-Mills
equation (\ref{YM-equation}) with (\ref{U^dagger}), we obtain
\begin{eqnarray}
 S_{YM} &=& \mbox{tr} \sum_{i,j,k} \, ( U_{kj} \, U_{ji}
          \, U_{ij} \, U_{jk} - U_{kj} \, U_{ji} \, U_{ik}
          - U_{ki} \, U_{ij} \, U_{jk} + U_{ki} \, U_{ik} )
                                      \label{yma}
\end{eqnarray}
and
\begin{eqnarray}
 \sum_k \, (F_{ikj} - \delta_{ij} \, F_{iki} - U_{ik} \, F_{kij}
       - F_{ijk} \, U_{kj}) = 0
\end{eqnarray}
respectively.
\vskip.3cm
\noindent
{\em Example.} For ${\cal M} = \Ir_2 = \lbrace 0,1 \rbrace$ with the
universal differential algebra one finds
\begin{eqnarray}
   F &=& e_{010} \, ( U_{01} U_{10} - {\bf 1})
       + e_{101} \, ( U_{10} U_{01} - {\bf 1})   \nonumber \\
     &=& e_{010} \, ( U_{10}^\dagger U_{10} - {\bf 1})
       + e_{101} \, ( U_{10} U_{10}^\dagger - {\bf 1})
\end{eqnarray}
and
\begin{eqnarray}
  S_{YM} = 2 \, \mbox{tr} \, ( U_{10}^\dagger U_{10} - {\bf 1})^2
\end{eqnarray}
which has the form of a Higgs potential (cf \cite{Conn+Lott90}).
                                           \hfill  {\Large $\Box$}

\vskip.3cm

With $U$ we also have $\check{U} := \sum_{i,j} e_{ij} \,
U_{ji}^{-1}$ as a transport operator (more generally, this
is the case for a differential algebra with a `symmetric' graph).
Hence, there is another connection,
\begin{eqnarray}
    \check{A} := \sum_{i,j}  e_{ij} \, ( U_{ji}^{-1} - {\bf 1}) \; .
\end{eqnarray}
For the corresponding covariant exterior derivatives
$\check{D} \Psi := d\Psi + \check{A} \, \Psi$ and $\check{D} \alpha
:= d \alpha - \alpha \, \check{A}$ one finds
\begin{eqnarray}
 \check{D}\Psi = \sum_{i,j} e_{ij} \, \check{\nabla}_j \Psi(i)
 \quad , \quad
 \check{\nabla}_j \Psi(i) := U_{ji}^{-1} \, \Psi(j) - \Psi(i)    \\
 \check{D}\alpha = \sum_{i,j} e_{ij} \, \check{\nabla}_j \alpha(i)
                   U_{ji}^{-1}       \quad , \quad
 \check{\nabla}_j \alpha(i) := \alpha(j) \, U_{ji} - \alpha(i)  \; .
\end{eqnarray}
There are thus two different parallel transports
between any two points. This suggests to look for field
configurations where the two covariant derivatives associated with $U$
and $\check{U}$ coincide. The condition
\begin{eqnarray}
   (D\Psi)^\dagger = \check{D}(\Psi^\dagger)
\end{eqnarray}
is equivalent to
\begin{eqnarray}
              A^\dagger = - \check{A}     \label{cond2}
\end{eqnarray}
and leads to $F^\dagger = \check{F}$. Furthermore,
(\ref{cond2}) implies
\begin{eqnarray}
        U_{ij}^\dagger = U_{ij}^{-1}  \; .    \label{U-inv}
\end{eqnarray}
Using (\ref{U^dagger}), this yields
\begin{eqnarray}
         U_{ij}^{-1} = U_{ji}  \; . \label{U-extra}
\end{eqnarray}
As a consequence, we have $F_{iji} = 0$. The condition (\ref{cond2})
thus eliminates the gauge field in Connes' 2-point space
model \cite{Conn+Lott90} (see also the example given above).
The Yang-Mills equation is now reduced to
\begin{eqnarray}
       \sum_k \, ( U_{ij} - U_{ik} \, U_{kj} ) = 0  \; .
\end{eqnarray}
If $\cal M$ is a finite set of $N$ elements, the last equation can be
rewritten as
\begin{eqnarray}
  U_{ij} = {1\over N}   \, \sum_k \, U_{ik} \, U_{kj}
         = {1\over N-2} \, \sum_{k \neq i,j} \, U_{ik} \, U_{kj}
                                     \label{yme}
\end{eqnarray}
where the last equality assumes $N \neq 2$.

%\vskip.6cm

\unitlength1.cm
\begin{picture}(8.,4.)(-7.5,-1.)
\thinlines
\put(-1.,0.) {\circle*{0.2}}
\put(1.,0.) {\circle*{0.2}}
\put(0.,1.) {\circle*{0.2}}
\put(0.,2.) {\circle*{0.2}}
\put(1.,0.) {\vector(-1,0){1.9}}
\put(1.,0.) {\vector(-1,1){0.9}}
\put(1.,0.) {\vector(-1,2){0.9}}
\put(0.,1.) {\vector(-1,-1){0.9}}
\put(0.,2.) {\vector(-1,-2){0.9}}
\put(-1.4,-0.2) {$i$}
\put(1.2,-0.2) {$j$}
\put(-0.1,0.55) {$k$}
\put(-0.1,2.2) {$\ell$}
\put(-0.05,1.3) {$\vdots$}
\put(-.3,-.6) {$U_{ij}$}
\end{picture}

\centerline{
\begin{minipage}[t]{9cm}
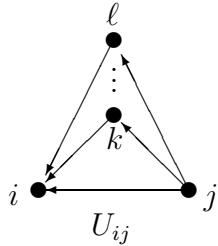
\centerline{\bf Fig.7}
\vskip.1cm \noindent
\small
\centerline{An illustration of equation (\ref{yme}).}
\end{minipage}    }

\vskip.4cm
\noindent
(\ref{yme}) means that the parallel transport operator from $j$ to
$i$ equals the average of all parallel transports via some other
point $k \neq i,j$ (see Fig.7).
Evaluation of the Yang-Mills action (\ref{yma}) with the condition
(\ref{cond2}) leads to the expression
\begin{eqnarray}
 S_{YM}' = 2 \, \sum_{i,j,k} \, \mbox{tr} \,
           (1 - U_{ij} \, U_{jk} \, U_{ki})
\end{eqnarray}
which contains a sum over all parallel transport loops with three
vertices (cf Fig.6). Note, however, that the above reduced Yang-Mills
equations are not obtained by variation of this action with respect
to $U_{ij}$ as a consequence of the constraint (\ref{cond2}).

\section{Lattice calculus}
\setcounter{equation}{0}
In this section we choose ${\cal M} = \Ir^n = \{ a=(a^\mu)| \;
\mu=1,\ldots,n; \; a^\mu \in \Ir \}$ and consider the reduction of the
universal differential algebra obtained by imposing the relations
\begin{eqnarray}
 e_{ab} \neq 0  \qquad \Leftrightarrow \qquad  b = a + \hat{\mu}
 \; \mbox{ for some } \mu                           \label{lrl}
\end{eqnarray}
where $\hat{\mu} := (\delta^\nu_\mu) \in {\cal M}$.
The corresponding graph is an oriented lattice in $n$ dimensions
(a finite part of it is drawn in Fig.8).

\unitlength1.cm
\begin{picture}(6,6.5)(-5.,-.5)
\thicklines
%\linethickness{0.3mm}
%
\multiput(0,0)(1,0){6}{\circle*{0.15}}
\multiput(0,1)(1,0){6}{\circle*{0.15}}
\multiput(0,2)(1,0){6}{\circle*{0.15}}
\multiput(0,3)(1,0){6}{\circle*{0.15}}
\multiput(0,4)(1,0){6}{\circle*{0.15}}
\multiput(0,5)(1,0){6}{\circle*{0.15}}
\multiput(0,0)(1,0){5}{\vector(1,0){0.9}}
\multiput(0,1)(1,0){5}{\vector(1,0){0.9}}
\multiput(0,2)(1,0){5}{\vector(1,0){0.9}}
\multiput(0,3)(1,0){5}{\vector(1,0){0.9}}
\multiput(0,4)(1,0){5}{\vector(1,0){0.9}}
\multiput(0,5)(1,0){5}{\vector(1,0){0.9}}
\multiput(0,0)(0,1){5}{\vector(0,1){0.9}}
\multiput(1,0)(0,1){5}{\vector(0,1){0.9}}
\multiput(2,0)(0,1){5}{\vector(0,1){0.9}}
\multiput(3,0)(0,1){5}{\vector(0,1){0.9}}
\multiput(4,0)(0,1){5}{\vector(0,1){0.9}}
\multiput(5,0)(0,1){5}{\vector(0,1){0.9}}
\end{picture}

\centerline{
\begin{minipage}[t]{9cm}
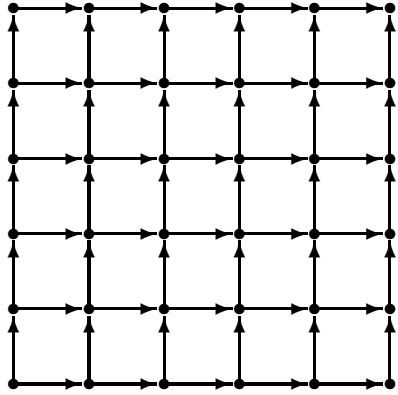
\centerline{\bf Fig.8}
\vskip.1cm \noindent
\small
A finite part of the oriented lattice graph which determines
the differential calculus underlying usual lattice theories.
\end{minipage}    }

\subsection{Differential calculus on the oriented lattice}
In the following we will use the notation
$e^\mu_a := e_{a,a+\hat{\mu}}$ and
\begin{eqnarray}
 e^{\mu_1 \ldots \mu_r}_a := e^{\mu_1}_a \, e^{\mu_2 \ldots
  \mu_r}_{a + \hat{\mu}_1}    \; .
\end{eqnarray}
In particular,
$e^{\mu \nu}_a = e_{a,a+\hat{\mu},a+\hat{\mu}+\hat{\nu}}
= e_{a,a+\hat{\mu}} \, e_{a+\hat{\mu}, a+\hat{\mu}+\hat{\nu}} $.
It also turns out to be convenient to introduce
\begin{eqnarray}
                   e^\mu := \sum_a e^\mu_a
\end{eqnarray}
which satisfies $e^\mu_a = e_a \, e^\mu$ and, more generally,
\begin{eqnarray}
  e^{\mu_1 \ldots \mu_r}_a = e_a \; e^{\mu_1} \cdots e^{\mu_r} \; .
\end{eqnarray}
\vskip.2cm
\noindent
Acting with $d$ on $e_{a,a+\hat{\mu}+\hat{\nu}}=0$ and using
(\ref{de_ij}), we obtain
\begin{eqnarray}
        e^{\mu\nu}_a + e^{\nu\mu}_a = 0           \label{l2r}
\end{eqnarray}
and therefore
\begin{eqnarray}             \label{ph-l2r}
        e^\mu \, e^\nu = - e^\nu \, e^\mu \; .
\end{eqnarray}
Using (\ref{drf}) we find
\begin{eqnarray}            \label{ph-e}
 de_a = \sum_\mu [ e_{a-\hat{\mu}} - e_a ] \, e^\mu  \quad , \quad
 d e^\mu_a = \sum_\nu [ e_{a-\hat{\nu}} - e_a ] \, e^\nu \,e^\mu
                         \; .
\end{eqnarray}
This leads to
\begin{eqnarray}
                   de^\mu = 0  \; .              \label{ph-m}
\end{eqnarray}
(\ref{ph-e}), (\ref{ph-m}) and the Leibniz rule allow us to calculate
$d \omega$ for any form $\omega$.
Any $f \in {\cal A}$ can be written as a function of
\begin{eqnarray}
              x^\mu := \ell \, \sum_a a^\mu \, e_a
\end{eqnarray}
(where $\ell$ is a positive constant) since
\begin{eqnarray}
   f = \sum_a e_a \, f(\ell a) = \sum_a e_a f(x) = f(x)
\end{eqnarray}
using $e_a \, x^\mu = \ell \, e_a \, a^\mu$ and $\sum_a e_a = \idty$.
The differential of a function $f$ is then given by
\begin{eqnarray}
 df &=& \sum_{\mu,a} e^\mu_a \, \lbrack f(\ell (a + \hat{\mu}))
        - f(\ell a)  \rbrack
     = \sum_{\mu,a} \lbrack f(x+\ell \hat{\mu}) - f(x) \rbrack \;
       e^\mu_a             \nonumber \\
    &=& \sum_\mu  (\partial_{+\mu}f)(x) \; dx^\mu
\end{eqnarray}
(where the expression $x + \ell \, \hat{\mu}$ should be read as $x +
\ell \, \hat{\mu} \, \idty$). We have introduced
\begin{eqnarray}
   (\partial_{\pm \mu}f)(x) := \pm {1 \over \ell} \,
    \lbrack f(x \pm \ell \hat{\mu}) - f(x) \rbrack
\end{eqnarray}
and
\begin{eqnarray}
          dx^\mu = \ell \, e^\mu    \; .
\end{eqnarray}
Furthermore, we obtain
\begin{eqnarray}
 dx^\mu \, f(x) & = & \ell \, \sum_a e^\mu_a \, f(x)
 = \ell \, \sum_a e^\mu_a \, f(\ell (a+\hat{\mu}))
 = \ell \, \sum_a f(\ell (a+\hat{\mu})) \, e^\mu_a
   \nonumber \\
 & = & \ell \, f(x + \ell \hat{\mu}) \, \sum_a e^\mu_a
   = f(x+\ell \hat{\mu}) \, dx^\mu                  \label{f-dx-latt}
\end{eqnarray}
which shows that we are dealing with the differential calculus of
\cite{DMH92,DMHS93} which was demonstrated to underly usual lattice
theories. The 1-forms $dx^\mu$ constitute a basis of
$\Omega^1({\cal A})$ as a left (or right) $\cal A$-module.
In accordance with previous results \cite{DMHS93} we obtain
\begin{eqnarray}
       dx^\mu \, dx^\nu + dx^\nu \, dx^\mu = 0
\end{eqnarray}
from (\ref{ph-l2r}). More generally,
$ dx^{\mu_1} \cdots dx^{\mu_r} = \ell^r \, e^{\mu_1} \cdots e^{\mu_r}$
is totally antisymmetric. For an arbitrary $r$-form
\begin{eqnarray}
 \phi = { 1 \over r!} \, \sum_{a, \mu_1, \ldots, \mu_r} \ell^r\,
        e^{\mu_1 \ldots \mu_r}_a \, \phi_{\mu_1 \ldots \mu_r}(\ell a)
      = { 1 \over r!} \, \sum_{\mu_1, \ldots, \mu_r} \phi_{\mu_1 \ldots
        \mu_r}(x) \, dx^{\mu_1} \cdots dx^{\mu_r}
\end{eqnarray}
one finds
\begin{eqnarray}
 d\phi  =  {1\over r!} \, \sum_{\mu, \mu_1, \ldots, \mu_r}
           (\partial_{+\mu} \phi_{\mu_1 \ldots \mu_r})(x) \;
           dx^\mu \, dx^{\mu_1} \cdots dx^{\mu_r}
\end{eqnarray}
using the rules of differentiation. With the help of (\ref{f-dx-latt}), the
differential of $\phi$ can also be expressed as
\begin{eqnarray}
       d \phi = \lbrack u , \phi \, \rbrace           \label{lud}
\end{eqnarray}
with the graded commutator on the rhs and the 1-form
\begin{eqnarray}
                  u := {1 \over \ell} \, \sum_\mu dx^\mu
\end{eqnarray}
which satisfies $u^2 = 0$.
\vskip.3cm

Next we introduce an inner product of forms.
Taking account of the identities (\ref{l2r}), we set
\begin{eqnarray}
 (e^{\mu_1 \ldots \mu_r}_a , e^{\nu_1 \ldots \nu_s}_b) :=
 \ell^{-2r} \,  \delta_{rs} \, \delta_{a,b} \,
 \delta^{\nu_1 \ldots \nu_r}_{\mu_1 \ldots\mu_r}
\label{lip}
\end{eqnarray}
where
\begin{eqnarray}
 \delta^{\nu_1 \ldots \nu_r}_{\mu_1 \ldots \mu_r} :=
    \delta^{\nu_1}_{[\mu_1} \cdots \delta^{\nu_r}_{\mu_r]}
  = \sum_{k=1}^r (-1)^{k+1} \, \delta^{\nu_1}_{\mu_k} \,
    \delta^{\nu_2 \ldots \ldots \ldots \nu_r}_{\mu_1 \ldots
    \widehat{\mu_k} \ldots \mu_r}     \; .
\end{eqnarray}
This is compatible with (\ref{innerpr}) since the rhs vanishes for
$a+ \hat{\mu}_1 + \ldots + \hat{\mu}_r \neq b + \hat{\nu}_1 +
\ldots + \hat{\nu}_r$. In particular, we get $(f,h) = \sum_a
\overline{f}(\ell a) \, h(\ell a)$ and
\begin{eqnarray}
 (\psi , \phi) = {1\over r!} \, \sum_{a,\mu_1, \ldots, \mu_r}
 \overline{\psi}_{\mu_1 \ldots \mu_r}(\ell a) \;
           \phi_{\mu_1 \ldots \mu_r}(\ell a)
\end{eqnarray}
for $r$-forms $\psi, \phi$.
\vskip.2cm

An adjoint $d^\ast$ of the exterior derivative $d$ can now be
introduced as in section II. A simple calculation shows that
\begin{eqnarray}
 d^\ast e^{\mu_1 \ldots \mu_r}(a) = \ell^{-2} \sum_{k=1}^r (-1)^{k+1}
 \, [e_{a+\hat{\mu}_k}-e_a] \, e^{\mu_1} \cdots \widehat{e^{\mu_k}}
 \cdots e^{\mu_r}                                       \label{ldast}
\end{eqnarray}
and
\begin{eqnarray}
 d^\ast\phi   =  - {1 \over (r-1)!} \, \sum_{\mu_1 \ldots,\mu_r}
      \partial_{-\mu_1} \phi_{\mu_1 \ldots \mu_r}(x) \,
      dx^{\mu_2} \cdots dx^{\mu_r}   \; .
\end{eqnarray}
\vskip.3cm
\noindent
{\em Remark.} From (\ref{ldast}) and (\ref{lip}) we find
\begin{eqnarray}
 (e^{\mu_1 \ldots \mu_r}_a , d^\ast e^{\nu_1 \ldots \nu_{r+1}}_b) =
 (-1)^{r+1} \, \ell^{-2(r+1)} \, I[(a;\mu_1 \ldots \mu_r) ,
 (b;\nu_1 \ldots \nu_{r+1})]
\end{eqnarray}
where
\begin{eqnarray}
  I[(a;\mu_1 \ldots \mu_r) , (b;\nu_1 \ldots \nu_{r+1})] =
 \sum_{k=1}^{r+1} (-1)^{r+1-k} \, (\delta_{a,b} - \delta_{a,b
   + \hat{\nu}_k}) \, \delta^{\nu_1 \ldots \widehat{\nu_k} \ldots
   \nu_{r+1}}_{\mu_1 \ldots \ldots \ldots \mu_r}
\end{eqnarray}
is the {\em incidence number} of the two cells
$(a;\mu_1 \ldots \mu_r)$ and $(b;\nu_1 \ldots \nu_{r+1})$.
This relates our formalism to others used in lattice theories
(cf \cite{Rabi81}, for example).  \\
\hspace*{1cm}                             \hfill  {\large $\Box$}

\vskip.3cm
\noindent
For the $\bullet$-products (cf section II) defined with respect
to the inner product introduced above one can prove the relations
\begin{eqnarray}
 (dx^\mu)^\natural \bullet (dx^{\mu_1} \cdots dx^{\mu_r}) &=&
 \sum_{k=1}^r (-1)^{k+1} \, \delta^\mu_{\mu_k} \, dx^{\mu_1} \cdots
 \widehat{dx^{\mu_k}} \cdots dx^{\mu_r}    \nonumber \\
 &=& (-1)^{r+1} \,
 (dx^{\mu_1} \cdots dx^{\mu_r}) \bullet (dx^\mu)^\natural \; .
                          \label{dx-bull}
\end{eqnarray}
Together with the general formulae given in section II, this allows
us to evaluate any expression involving a $\bullet$.

\vskip.3cm
\noindent
Let us introduce the `volume form'
\begin{eqnarray}
\epsilon :=  {1\over n!}\sum_{\mu_1,\ldots,\mu_n}\epsilon_{\mu_1
 \ldots \mu_n} \, dx^{\mu_1} \cdots dx^{\mu_n}
          = {\ell^n\over n!} \, \sum_{a,\mu_1,\ldots,\mu_n}
      \epsilon_{\mu_1 \ldots \mu_n} \, e^{\mu_1 \ldots \mu_n}_a
\end{eqnarray}
where $\epsilon_{\mu_1 \ldots \mu_n}$ is totally antisymmetric with
$\epsilon_{1 \ldots n} = 1$. Obviously, $d\epsilon=0$ and
$d^\ast\epsilon=0$. We can now define a Hodge star-operator on
differential forms as follows,
\begin{eqnarray}
   \star \, \psi := \psi^\natural \bullet \epsilon   \; .
\end{eqnarray}
Using (\ref{ph-i}), we obtain
\begin{eqnarray}
    \star \, (\psi f) = \overline{f} \, \star \psi   \; .
\end{eqnarray}
An application of (\ref{f-dx-latt}) leads to
\begin{eqnarray}
    \star \, (f \, \psi) = (\star \psi) \, \overline{f}(x-\ell
                           \hat{\epsilon})
\end{eqnarray}
where $\hat{\epsilon}:=\hat{1} + \ldots + \hat{n}$. The usual formula
\begin{eqnarray}
 \star(dx^{\mu_1} \cdots dx^{\mu_r}) = {1\over(n-r)!} \,
 \sum_{\mu_{r+1}, \ldots, \mu_n} \epsilon_{\mu_1 \ldots \mu_n} \,
 dx^{\mu_{r+1}} \cdots dx^{\mu_n}
\end{eqnarray}
holds as can be shown with the help of (\ref{2bull}) and
(\ref{dx-bull}). It can be used to show that
\begin{eqnarray}
 \star \, \star [\psi_r(x)] = (-1)^{r(n-r)} \, \psi_r(x-\ell
                              \hat{\epsilon})   \; .
\end{eqnarray}
A simple consequence of our definitions is
\begin{eqnarray}
 \epsilon \bullet \psi_r(x)^\natural = (-1)^{r(n+1)} \,
  \star \psi_r (x+\ell \hat{\epsilon})  \; .
\end{eqnarray}
With the help of this relation one finds
\begin{eqnarray}
   (\star \psi , \star \omega) = (\omega , \psi)  \; .
\end{eqnarray}
Furthermore, we have
\begin{eqnarray}
 d^\ast \psi_r(x) = -(-1)^{n(r+1)} \, \star \, d \, \star
    \psi_r(x+\ell\hat{\epsilon})   \; .
\end{eqnarray}
It is natural to introduce the following notation,
\begin{eqnarray}
     \int \omega_n := (\epsilon,\omega_n)  \; .
\end{eqnarray}

\subsection{Gauge theory on a lattice}
A connection 1-form can be written as $A = \sum_\mu  A_\mu(x) \, dx^\mu$.
Again, instead of $A$ we consider
\begin{eqnarray}
      U_\mu(x) := {\bf 1} + \ell A_\mu(x) \; .
\end{eqnarray}
The transformation law (\ref{A-transf}) for $A$ then leads to
\begin{eqnarray}
    U'_\mu(x) = G(x) \, U_\mu(x) \, G(x+\ell\hat{\mu})^{-1}  \; .
\end{eqnarray}
For the exterior covariant derivatives (\ref{DPsi}) we obtain
\begin{eqnarray}
 D \Psi = \sum_\mu \, \nabla_\mu \Psi(x) \, dx^\mu  \quad , \quad
 \nabla_\mu \Psi(x) := {1 \over \ell} \, [ U_\mu(x) \,
   \Psi(x+\ell \hat{\mu})-\Psi(x) ]  \quad    \\
 D \alpha = \sum_\mu \, \nabla_\mu \alpha(x) \, U_\mu(x) \, dx^\mu
    \quad , \quad
 \nabla_\mu \alpha(x) := {1 \over \ell} \, [ \alpha(x+\ell \hat{\mu})
    \, U_\mu(x)^{-1} - \alpha(x) ]    \; .
\end{eqnarray}
Using (\ref{lud}) and $u^2 = 0$, we find
\begin{eqnarray}
       F = dA + A^2 = u \, A + A \, u + A^2 = U^2
\end{eqnarray}
for the curvature of the connection $A$. Here we have introduced
\begin{eqnarray}
       U := u + A = {1 \over \ell} \, \sum_\mu U_\mu \, dx^\mu \; .
\end{eqnarray}
Further evaluation of the expression for $F$ leads to
\begin{eqnarray}
 F = {1 \over 2 \ell^2} \, \sum_{\mu,\nu} \, \lbrack U_\mu(x) \,
     U_\nu(x+\ell\hat{\mu}) -  U_\nu(x) \,
     U_\mu(x+\ell\hat{\nu}) \rbrack \, dx^\mu \, dx^\nu   \; .
\end{eqnarray}
Imposing the compatibility condition
\begin{eqnarray}
  [\nabla_\mu \Psi(x)]^\dagger = \nabla_\mu[\Psi(x)]^\dagger
\end{eqnarray}
for the covariant derivative with a conjugation
leads to the unitarity condition $U_\mu(x)^\dagger = U_\mu(x)^{-1}$.
The Yang-Mills action $S_{YM}=\mbox{tr} \, (F,F)$ is now turned
into the Wilson action
\begin{eqnarray}
 S_{YM} = {1\over \ell^4} \, \sum_{a,\mu,\nu} \, \mbox{tr}
    \lbrack {\bf 1} - U_\mu(\ell a) \, U_\nu(\ell(a+\hat{\mu})) \,
    U_\mu(\ell(a+\hat{\nu}))^\dagger \, U_\nu(\ell a)^\dagger
    \rbrack                 \label{Wilson}
\end{eqnarray}
of lattice gauge theory. We also have
\begin{eqnarray}
S_{YM} = \mbox{tr} (F,F) = \mbox{tr} (\star F, \star F) =
         \mbox{tr} (\epsilon, F \star F) = \mbox{tr} \int F
         \star F    \; .
\end{eqnarray}
The Yang-Mills equations are again obtained in the form
(\ref{YM-equation}). Evaluation using (\ref{bullet-e1}) and
(\ref{bullet-e2}) leads to the lattice Yang-Mills equations
\begin{eqnarray}
   U_\mu(x) &=& {1\over 2n} \, \sum_\nu \lbrack U_\nu(x) \,
      U_\mu(x+\ell\hat{\nu}) \, U_\nu(x+\ell\hat{\mu})^\dagger
      \nonumber \\
 & & + U_\nu(x-\ell\hat{\nu})^\dagger \, U_\mu(x-\ell\hat{\nu}) \,
           U_\nu(x+\ell(\hat{\mu}-\hat{\nu})) \rbrack
\end{eqnarray}
which have a simple geometric meaning on the lattice.
$U_\mu(x)$ must be the average of the parallel transports
along all neighbouring paths.

\section{The symmetric lattice.}
\setcounter{equation}{0}
The lattice considered in the previous section had an orientation
arbitrarily assigned to it (the arrows point in the direction of
increasing values of the coordinates $x^\mu$).
In this section we consider a `symmetric lattice', i.e. a lattice
without distinguished directions. It corresponds to a `symmetric
reduction' (of the universal differential algebra on $\Ir^n$) as
defined in section III. Some of its features were anticipated in
example 5 of section III.
\vskip.3cm

The differential calculus associated with the symmetric lattice
turns out to be a kind of discrete version of a
`noncommutative differential calculus' on manifolds which has been
studied recently \cite{DMH92-grav,DMH93-stoch,MH+Reut93}.
\vskip.3cm

Again, we choose ${\cal M}=\Ir^n$ and use the same notation as in the
previous section. The reduction of the universal differential
algebra associated with a `symmetric' ($n$-dimensional) lattice
is determined by
\begin{eqnarray}
 e_{ab} \neq 0 \quad \Leftrightarrow \quad
 b = a + \hat{\mu} \quad \mbox{or} \quad b = a-\hat{\mu}
 \quad \mbox{for some } \mu
\end{eqnarray}
where $\hat{\mu} = (\delta^\nu_\mu)$ and $\mu = 1, \ldots, n$.

\subsection{Calculus on the symmetric lattice}
It is convenient to introduce a variable $\epsilon$ which takes values
in $\lbrace \pm 1 \rbrace$. Furthermore, we define $e^{\epsilon \mu}_a
:= e_{a , a + \epsilon \hat{\mu}}$ and, more generally,
\begin{eqnarray}
 e^{\epsilon_1 \mu_1 \ldots \epsilon_r \mu_r}_a := e^{\epsilon_1
 \mu_1}_a \; e^{\epsilon_2 \mu_2 \ldots \epsilon_r
 \mu_r}_{a+\epsilon_1 \hat{\mu}_1}  \; .
\end{eqnarray}
Acting with $d$ on the identity $e_{a , a+\epsilon \hat{\mu} + \epsilon'
\hat{\nu}} = 0$ for $\epsilon \, \hat{\mu} + \epsilon' \, \hat{\nu}
\neq 0$, we obtain
\begin{eqnarray}
 e^{\epsilon \mu \, \epsilon' \nu}_a + e^{\epsilon' \nu \,
 \epsilon \mu}_a = 0
 \qquad (\epsilon \, \hat{\mu} + \epsilon' \, \hat{\nu} \neq 0)  \; .
\end{eqnarray}
We supplement these relations with corresponding relations for
$\epsilon \, \hat{\mu} + \epsilon' \, \hat{\nu} = 0$, namely
\begin{eqnarray}
   e^{+ \mu \, - \mu}_a + e^{- \mu \, + \mu}_a = 0  \; .
                                            \label{slanti}
\end{eqnarray}
The general case will be discussed elsewhere. In (\ref{slanti})
we have simply written $\pm$ instead of $\pm 1$.
As a consequence of (\ref{slanti}), $e^{\epsilon_1 \mu_1 \ldots
\epsilon_r \mu_r}_a$ is totally antisymmetric in the (double-)
indices $\epsilon_i \mu_i$. Introducing
\begin{eqnarray}
    e^{\epsilon \mu} := \sum_a e^{\epsilon \mu}_a
\end{eqnarray}
we have
\begin{eqnarray}
  \sum_a e^{\epsilon_1 \mu_1 \ldots \epsilon_r \mu_r}_a
  = e^{\epsilon_1 \mu_1} \cdots e^{\epsilon_r \mu_r}
\end{eqnarray}
and
\begin{eqnarray}
 e_a^{\epsilon_1 \mu_1} \, e^{\epsilon_2 \mu_2}
 \cdots e^{\epsilon_r \mu_r} =
 e_a \, e^{\epsilon_1 \mu_1} \cdots e^{\epsilon_r \mu_r} \; .
\end{eqnarray}
As a consequence of these relations we find
\begin{eqnarray}
 e^{\epsilon \mu \, \epsilon' \nu} + e^{\epsilon' \nu \,
 \epsilon \mu} = 0
\end{eqnarray}
and the general differentiation rule (\ref{drf}) gives
\begin{eqnarray}
 d \lbrack e_a^{\epsilon_1 \mu_1} \, e^{\epsilon_2 \mu_2}
 \cdots e^{\epsilon_r \mu_r}
  \rbrack = \sum_{\epsilon,\mu} \lbrack e_{a - \epsilon \hat{\mu}}
  - e_a \rbrack \; e^{\epsilon \mu} \, e^{\epsilon_1 \mu_1}
  \cdots e^{\epsilon_r \mu_r}                      \label{sld}
\end{eqnarray}
which, in particular, implies $d e^{\epsilon\mu} = 0$.
As in the previous section we introduce
\begin{eqnarray}
    x^\mu := \ell \, \sum_a a^\mu \, e_a     \; .
\end{eqnarray}
Every $f \in {\cal A}$ can be regarded as a
function of $x^\mu$ (cf section V). Using (\ref{sld}) we obtain
\begin{eqnarray}
 dx^\mu = \ell \, (e^{+\mu} - e^{-\mu}) = \ell \, \sum_{\epsilon}
          \epsilon \, e^{\epsilon \mu}
\label{sldx}\end{eqnarray}
and thus
\begin{eqnarray}
 e^{\epsilon \mu} = {\epsilon \over 2 \ell}  \, dx^\mu +
                    {1 \over 2 \beta} \, \tau^\mu
\end{eqnarray}
with the 1-forms
\begin{eqnarray}
  \tau^\mu := \beta \, (e^{+\mu} + e^{-\mu}) = \beta \,
              \sum_{\epsilon} e^{\epsilon \mu}       \label{sltau}
\end{eqnarray}
where $\beta \neq 0$ is a real constant. The 1-forms $\tau^\mu$
satisfy $d \tau^\mu = 0$. Together with $dx^\mu$ they form a basis
of $\Omega^1({\cal A})$ as a left (or right) $\cal A$-module.
The differential of a function $f(x)$
can now be written as follows,
\begin{eqnarray}
 df = \ell \, \sum_{\epsilon,\mu} \epsilon \,
      \partial_{\epsilon \mu} f \, e^{\epsilon \mu}
    = \sum_\mu (\bar{\partial}_\mu f \, dx^\mu + {\kappa\over2} \,
       \Delta_\mu f \, \tau^\mu )
\end{eqnarray}
where $\kappa:=\ell^2/\beta$ and we have introduced the operators
\begin{eqnarray}
 \partial_{\epsilon\mu} f & := & {\epsilon \over \ell} \, \left( f(x
    + \epsilon \, \ell \, \hat{\mu}) - f(x) \right)             \\
 \bar{\partial}_\mu f & := & {1\over 2} \, (\partial_{+\mu} f +
    \partial_{-\mu} f) = {1 \over2 \ell} \, \left\lbrack f(x
    + \ell \, \hat{\mu}) - f(x - \ell \, \hat{\mu}) \right\rbrack  \\
 \Delta_\mu f & := & \partial_{+\mu} \, \partial_{-\mu} f =
    {1 \over \ell} \, (\partial_{+\mu} f - \partial_{-\mu} f)
                                                      \nonumber  \\
 & = & {1 \over \ell^2} \, \left\lbrack f(x + \ell \, \hat{\mu})
    + f(x - \ell \, \hat{\mu}) - 2 \, f(x) \right\rbrack  \; .
\end{eqnarray}
For the commutation relations between functions and 1-forms we find
\begin{eqnarray}            \label{epsmu-shift}
  e^{\epsilon \mu} \, f(x) =  f(x + \epsilon \, \ell \, \hat{\mu})
                              \, e^{\epsilon \mu}
\end{eqnarray}
and
\begin{eqnarray}
 \lbrack dx^\mu , f(x) \rbrack & = & {\kappa \beta \over 2} \,
     \Delta_\mu f(x) \, dx^\mu + \kappa \, \bar{\partial}_\mu f(x)
     \, \tau^\mu                         \label{f,dx-comm}       \\
 \lbrack \tau^\mu , f(x) \rbrack & = &  \beta \,
     \bar{\partial}_\mu f(x) \, dx^\mu + {\kappa \beta \over 2} \,
     \Delta_\mu f(x) \, \tau^\mu \; .         \label{f,tau-comm}
\end{eqnarray}
\vskip.3cm

Let us take a look at the continuum limit where
$\ell \to 0$ and $\beta \to 0$, but $\kappa=$const. Under the
additional assumption that $ \tau^\mu \rightarrow \tau $,
one (formally) obtains from (\ref{f,dx-comm}) and (\ref{f,tau-comm})
the commutation relations
\begin{eqnarray}
 \lbrack dx^\mu , f(x) \rbrack & = &
     \kappa \, \delta^{\mu\nu} \partial_\nu f(x) \, \tau
       \qquad   \mbox{(summation over $\nu$)}   \label{ndc1}    \\
 \lbrack \tau , f(x) \rbrack & = & 0            \label{ndc2}
\end{eqnarray}
with the metric tensor $\delta^{\mu\nu}$.
For the differential of a (differentiable) function we get
\begin{eqnarray}
 df = \partial_\mu f \, dx^\mu + {\kappa\over2} \, \Box f \, \tau
      \qquad   \mbox{(summation over $\mu$)}
\end{eqnarray}
in the continuum limit. Here $\Box := \sum_{\mu \nu} \delta^{\mu \nu}
 \, \partial_\mu \, \partial_\nu$ is the d'Alembertian of the metric
$\delta^{\mu\nu}$ and $\partial_\mu$ is the ordinary partial derivative
with respect to $x^\mu$. Differential calculi of the form
(\ref{ndc1}), (\ref{ndc2}) on manifolds have been investigated
recently. They are related to quantum theory \cite{DMH92-grav} and
stochastics \cite{DMH93-stoch} and show up in the classical limit
of (bicovariant) differential calculi on certain quantum groups
\cite{MH+Reut93}.
\vskip.3cm

Returning to the general case, we find from (\ref{slanti}) the 2-form
relations
\begin{eqnarray}
 dx^\mu \, dx^\nu + dx^\nu \, dx^\mu= 0 \quad , \quad
 dx^\mu \, \tau^\nu + \tau^\nu \, dx^\mu = 0 \quad , \quad
 \tau^\mu \, \tau^\nu + \tau^\nu \, \tau^\mu = 0  \; .
\end{eqnarray}
An $r$-form $\psi$ can be written in the following two ways,
\begin{eqnarray}
 \psi & = & {1\over r!} \,
 \sum_{\epsilon_1,\ldots , \epsilon_r \atop \mu_1,\ldots,\mu_r}
 \psi_{\epsilon_1 \mu_1 \cdots \epsilon_r \mu_r} \,
 e^{\epsilon_1 \mu_1} \cdots e^{\epsilon_r \mu_r}         \nonumber    \\
 & = & {1\over r!} \, \sum_{\mu_1,\ldots,\mu_r} \sum_{k=0}^r
       {r \atopwithdelims() k} \,\psi^{(k)}_{\mu_1 \cdots \mu_r} \,
       \tau^{\mu_1} \cdots \tau^{\mu_k} \, dx^{\mu_{k+1}} \cdots
       dx^{\mu_r} \; .
\end{eqnarray}
Using (\ref{sldx}) and (\ref{sltau}) we obtain
\begin{eqnarray}
 \psi_{\epsilon_1 \mu_1 \ldots \epsilon_r \mu_r} =
  \sum_{k=0}^r {\ell^{r-k} \beta^k \over k! \, (r-k)!} \, \sum_{\pi
  \in {\cal S}_r} \mbox{sgn} \pi \; \epsilon_{\pi(k+1)} \cdots
  \epsilon_{\pi(r)} \, \psi^{(k)}_{\mu_{\pi(1)} \ldots \mu_{\pi(r)}}
       \; .
\end{eqnarray}
Furthermore, we have
\begin{eqnarray}
             d \psi =  u \, \psi - (-1)^r \, \psi \, u
                    = \lbrack u , \psi \, \rbrace
\end{eqnarray}
with
\begin{eqnarray}                 \label{u}
  u :=  \sum_{\epsilon,\mu} e^{\epsilon \mu}
     = {1 \over \beta} \, \sum_\mu \tau^\mu
\end{eqnarray}
which satisfies $u^2=0$.
\vskip.3cm

An inner product is determined by
\begin{eqnarray}
 (e_a^{\epsilon_1 \mu_1} \cdots e^{\epsilon_r \mu_r} ,
 e_b^{\epsilon'_1 \nu_1} \cdots e^{\epsilon'_s \nu_s}) := \delta_{rs}
 \, (2\ell^2)^{-r} \, \delta_{a,b} \,
 \delta^{\epsilon'_1 \nu_1 \ldots \epsilon'_r
 \nu_r}_{\epsilon_1 \mu_1 \ldots \epsilon_r \mu_r}  \label{sl-ip}
\end{eqnarray}
and the usual rules (\ref{ip-rules}). The adjoint $d^\ast$ of $d$
with respect to this inner product then acts as follows,
\begin{eqnarray}
 d^\ast e^{\epsilon_1 \mu_1} \cdots e^{\epsilon_r \mu_r} =
 {1\over 2 \ell^2} \, \sum_{k=1}^r (-1)^{k+1} \,
 [e_{a+\epsilon_k \hat{\mu}_k} - e_a] \, e^{\epsilon_1 \mu_1} \cdots
 \widehat{e^{\epsilon_k \mu_k}} \cdots  e^{\epsilon_r \mu_r}   \; .
\end{eqnarray}
More generally, for an $r$-form $\psi$ we have
\begin{eqnarray}
 d^\ast \psi & = & - {1 \over 2 \ell} \, {1 \over (r-1)!} \,
 \sum_{\epsilon_1, \ldots, \epsilon_r \atop \mu_1,\ldots,\mu_r}
 \epsilon_1 \, \partial_{-\epsilon_1 \mu_1} \psi_{\epsilon_1 \mu_1
 \ldots \epsilon_r \mu_r} \, e^{\epsilon_2 \mu_2} \cdots
 e^{\epsilon_r\mu_r}     \nonumber \\
 & = & {1\over (r-1)!} \sum_{\mu_1,\ldots,\mu_r}
 \sum_{k=0}^r \left\lbrack {\beta \over 2} \, {r-1 \atopwithdelims()
 k-1} \, (\Delta_{\mu_1} \psi^{(k)}_{\mu_1 \ldots \mu_r}) \,
 \tau^{\mu_2} \cdots \tau^{\mu_k} \, dx^{\mu_{k+1}} \cdots dx^{\mu_r}
  \right .      \nonumber \\
 & & \left . + (-1)^{k+1} {r-1 \atopwithdelims() k} \,
 (\bar{\partial}_{\mu_{k+1}}\psi^{(k)}_{\mu_1 \ldots \mu_r}) \,
 \tau^{\mu_1} \cdots \tau^{\mu_k} \, dx^{\mu_{k+2}} \cdots dx^{\mu_r}
 \right\rbrack   \; .
\end{eqnarray}
For a 1-form $\rho=\sum_\mu ( \rho^{(0)}_\mu \,dx^\mu
+ \rho^{(1)}_\mu \, \tau^\mu)$ this reads
\begin{eqnarray}
 d^\ast \rho = - \sum_{\mu,\nu} \delta^{\mu\nu} \, (\bar{\partial}_\mu
 \rho^{(0)}_\nu - {\beta \over 2} \, \Delta_\mu \rho^{(1)}_\nu)
    \; .
\end{eqnarray}
If $\rho = df$, this implies
\begin{eqnarray}
 - d^\ast d f = \sum_{\mu,\nu} \delta^{\mu\nu} \, (\bar{\partial}_\mu
 \bar{\partial}_\nu f - {\ell^2 \over 2} \, \Delta_\mu \Delta_\nu f)
  = \sum_\mu  \Delta_\mu f
\end{eqnarray}
where we made use of the identity
\begin{eqnarray}
 \bar{\partial}_\mu^2 f - {\ell^2 \over 4} \, \Delta_\mu^2 f
  = \Delta_\mu f  \; .
\end{eqnarray}
In the continuum limit one thus obtains $-d^\ast d f = \delta^{\mu\nu}
\, \partial_\mu \partial_\nu f = \Box f$ (summation over $\mu$ and
$\nu$).
\vskip.3cm

The involution on $\Omega({\cal A})$ (induced by the identity on
$\cal M$) introduced in section II acts on the basic 1-forms as follows,
\begin{eqnarray}
 (e^{\epsilon \mu}_a)^\ast = - e^{-\epsilon \mu}_{a+\epsilon
                             \hat{\mu}}  \; .
\end{eqnarray}
This leads to
\begin{eqnarray}
 (e^{\epsilon\mu})^\ast = -e^{-\epsilon\mu} \quad , \quad
 (dx^\mu)^\ast = dx^\mu \quad , \quad (\tau^\mu)^\ast = -\tau^\mu
    \; .
\end{eqnarray}
The $\bullet$-products are again defined by (\ref{bullets}), now
with respect to the inner product (\ref{sl-ip}).
For their evaluation it is sufficient to know that
\begin{eqnarray}
 e^{\epsilon \mu} \bullet (e^{\epsilon_1 \mu_1} \cdots e^{\epsilon_r
 \mu_r}) = {1\over 2 \, \ell^2} \, \sum_{k=1}^r (-1)^{k+1} \,
 \, \delta^{\epsilon_k\mu_k}_{\epsilon\mu} \,
 e^{\epsilon_1 \mu_1} \cdots \widehat{e^{\epsilon_k \mu_k}} \cdots
 e^{\epsilon_r \mu_r}  \; .
\end{eqnarray}
In particular, one obtains
\begin{eqnarray}
\lefteqn{(\tau^\mu)^\natural \bullet (\tau^{\mu_1} \cdots \tau^{\mu_s}
 \, dx^{\mu_{s+1}} \cdots dx^{\mu_r})} \nonumber \\
 &  & = {\beta \over \kappa} \, \sum_{k=1}^{s} (-1)^{k+1} \,
 \delta^{\mu\mu_k} \,
 \tau^{\mu_1} \cdots \widehat{\tau^{\mu_k}} \cdots \tau^{\mu_s} \,
 dx^{\mu_{s+1}} \cdots dx^{\mu_r}                    \\
 \lefteqn{(dx^\mu)^\natural \bullet (\tau^{\mu_1} \cdots \tau^{\mu_s}
 \, dx^{\mu_{s+1}} \cdots dx^{\mu_r})}  \nonumber \\
&  & =\sum_{k=s+1}^{r} (-1)^{k+1} \, \delta^{\mu\mu_k} \,
 \tau^{\mu_1} \cdots \tau^{\mu_s} \, dx^{\mu_{s+1}} \cdots
 \widehat{dx^{\mu_k}} \cdots dx^{\mu_r}
\end{eqnarray}
and
\begin{eqnarray}
 (\tau^{\mu_1} \cdots \tau^{\mu_s} \, dx^{\mu_{s+1}} \cdots
 dx^{\mu_r}) \bullet (e^{\epsilon\mu})^\natural
 =  (-1)^{r+1} \, (e^{\epsilon \mu})^\natural \bullet (\tau^{\mu_1}
 \cdots \tau^{\mu_s} \, dx^{\mu_{s+1}} \cdots dx^{\mu_r})  \; .
\end{eqnarray}
Using these expressions one can show that
\begin{eqnarray}
 (\psi,\phi) = {1 \over r!} \, \sum_{a,\mu_1,\ldots,\mu_r} \sum_{k=0}^r
  {r \atopwithdelims() k} \left({\beta \over \kappa}\right)^k \,
  \overline{\psi}^{(k)}_{\mu_1 \cdots \mu_r}(\ell \, a)  \;
  \phi^{(k)}_{\mu_1 \cdots \mu_r}(\ell \, a)
                            \label{sl-innerprod}
\end{eqnarray}
for $r$-forms $\psi, \phi$.

\subsection{Gauge theory on the symmetric lattice}
A connection 1-form on the symmetric lattice can be expressed as
\begin{eqnarray}
 A =  \sum_{\epsilon,\mu} A_{\epsilon \mu} \, e^{\epsilon \mu}
   =  \sum_\mu (A^{(0)}_\mu \, dx^\mu + {\kappa\over2} \, A^{(1)}_\mu
      \, \tau^\mu)
\end{eqnarray}
where $A_{\epsilon \mu} = \epsilon \, \ell \, A^{(0)}_\mu +
(\ell^2/2) \, A^{(1)}_\mu$.
The transformation rule (\ref{A-transf}) for a connection 1-form
leads to
\begin{eqnarray}
 0 & = & \sum_\mu \left\lbrace \left[ \bar{\partial}_\mu G
       - G  \, A^{(0)}_\mu + A'{}^{(0)}_\mu \,(G + {\kappa\beta\over 2}
         \Delta_\mu G) + {\kappa\beta \over 2} \, A'{}^{(1)}_\mu
         \bar{\partial}_\mu G) \right]dx^\mu \right.   \nonumber \\
   &   & \, \quad + \left.  {\kappa\over2} \, \left[ \Delta_\mu G
         - G \, A^{(1)}_\mu + A'{}^{(1)}_\mu \, (G + {\kappa\beta \over 2} \,
         \Delta_\mu G) + 2 \,A'{}^{(0)}_\mu \,\bar{\partial}_\mu G
         \right] \tau^\mu \right\rbrace  \; .
\end{eqnarray}
Formally, the continuum limit $\ell\rightarrow 0$ of this equation
yields the familiar gauge transformation formula $\partial_\mu G =
G \, A^{(0)}_\mu - A'{}^{(0)}_\mu \, G$ for $A^{(0)}$ and in addition
\begin{eqnarray}
 \Box G = - 2 \, \sum_{\mu , \nu} \delta^{\mu\nu} \, A'{}^{(0)}_\nu \,
          \partial_\mu G + G \, A^{(1)} - A'{}^{(1)} \, G
\end{eqnarray}
where  $A^{(1)} := \sum_\mu \, A^{(1)}_\mu$.
\vskip.3cm

In terms of the basis $dx^\mu, \tau^\nu$, the curvature 2-form
$F = dA + A^2$ reads
\begin{eqnarray}
F & = & \sum_{\mu,\nu} \left\{ \left[ \bar{\partial}_\mu A^{(0)}_\nu +
  A^{(0)}_\mu A^{(0)}_\nu +{\kappa \beta \over 2} (A^{(0)}_\mu \,
  \Delta_\mu A^{(0)}_\nu + A^{(1)}_\mu \, \bar{\partial}_\mu
  A^{(0)}_\nu) \right] \, dx^\mu dx^\nu \right.  \nonumber \\
 & & + {\kappa \over 2} \, \left[ \Delta_\mu A^{(0)}_\nu
  - \bar{\partial}_\nu A^{(1)}_\nu - A^{(0)}_\nu A^{(1)}_\mu
  + A^{(1)}_\mu A^{(0)}_\nu + 2 A^{(0)}_\mu \, \bar{\partial}_\mu
  A^{(0)}_\nu  \right.                \nonumber \\
 & & \left. - {\kappa \beta \over 2} \, (A^{(0)}_\nu \Delta_\nu
  A^{(1)}_\mu -2 A^{(1)}_\mu \, \Delta_\mu A^{(0)}_\nu + A^{(1)}_\nu
  \, \bar{\partial}_\nu A^{(1)}_\mu) \right] \,
 \tau^\mu dx^\nu  \nonumber \\
 &   & \left.  + {\kappa^2 \over 4} \, \left[ \Delta_\mu A^{(1)}_\nu
  + 2A^{(0)}_\mu \, \bar{\partial}_\mu A^{(1)}_\mu
  + A^{(1)}_\mu A^{(1)}_\nu + {\kappa \beta \over 2} A^{(1)}_\mu
  \, \Delta_\mu A^{(1)}_\nu \right] \,
 \tau^\mu \tau^\nu \right\}    \nonumber    \\
 &=:& \sum_{\mu,\nu} \, [ {1 \over 2} \, F^{(0)}_{\mu \nu} \, dx^\mu \,
      dx^\nu + F^{(1)}_{\mu\nu} \, \tau^\mu \, dx^\nu + {1 \over 2}
      \, F^{(2)}_{\mu\nu} \, \tau^\mu \, \tau^\nu]   \; .
\end{eqnarray}
Evaluation of the Yang-Mills action with the help of
(\ref{sl-innerprod}) leads to
\begin{eqnarray}
 S_{YM} = \mbox{tr} \sum_{a,\mu,\nu}
 \left[ {1 \over 2} \, F^{(0)}_{\mu\nu}{}^\dagger \, F^{(0)}_{\mu\nu}
 + {\beta \over \kappa} \, F^{(1)}_{\mu\nu}{}^\dagger \,
 F^{(1)}_{\mu\nu} + {1 \over 2} \, \left({\beta \over \kappa} \right)^2
 F^{(2)}_{\mu\nu}{}^\dagger \, F^{(2)}_{\mu\nu} \right]
\end{eqnarray}
where the function in square brackets has to be taken at $\ell \, a$.
Obviously, because of the factors $\beta/\kappa$
the ordinary Yang-Mills action for $A^{(0)}_\mu$ is obtained
in the limit $\ell \to 0, \beta \to 0$ (with $\kappa$ fixed).
\vskip.3cm

Again, we introduce
\begin{eqnarray}
  U_{\epsilon \mu} = {\bf 1} + A_{\epsilon \mu} = {\bf 1} + \epsilon
  \, \ell \, A^{(0)}_\mu + {\ell^2 \over2} \, A^{(1)}_\mu
\end{eqnarray}
which transforms as follows,
\begin{eqnarray}
  U'_{\epsilon \mu}(x) = G(x) \, U_{\epsilon\mu}(x) \,
       G(x + \epsilon \ell \hat{\mu})^\dagger
\end{eqnarray}
(note that $G^\dagger = G^{-1}$). Using (\ref{epsmu-shift}) this
implies that
\begin{eqnarray}
   E^{\epsilon \mu} := U_{\epsilon \mu} \, e^{\epsilon \mu}
\end{eqnarray}
transform covariantly under a gauge transformation, i.e.,
$G \, E^{\epsilon \mu} \, G^\dagger = U'_{\epsilon \mu} \,
e^{\epsilon \mu} = E'{}^{\epsilon \mu}$.
Also covariant are the 1-forms
\begin{eqnarray}
 Dx^\mu := ({\bf 1} + {\kappa \beta \over 2} \, A^{(1)}_\mu) \,
           dx^\mu + \kappa A^{(0)}_\mu \, \tau^\mu
         = U \, x^\mu - x^\mu \, U
\end{eqnarray}
where
\begin{eqnarray}
 U := u + A = \sum_{\epsilon, \mu} U_{\epsilon \mu} \, e^{\epsilon \mu}
    = \sum_{\epsilon, \mu} E^{\epsilon \mu}
\end{eqnarray}
with $u$ defined in (\ref{u}). Together with $E^{\epsilon \mu}$ the
$Dx^\mu$ constitute a basis of the space of 1-forms (as a left or right
$\cal A$-module) and allow us to read off covariant components from
covariant differential forms.
\vskip.3cm

For the covariant exterior derivatives (\ref{DPsi}) we find
\begin{eqnarray}
 D \Psi = U \, \Psi - \Psi \, u   \quad , \quad
 D \alpha = u \, \alpha - \alpha \, U       \; .
\end{eqnarray}
\vskip.3cm

In the following we constrain $U$ with the conditions
\begin{eqnarray}
 U_{-\epsilon \mu}(x + \epsilon \, \ell \, \hat{\mu})
      = U_{\epsilon \mu}(x)^\dagger = U_{\epsilon \mu}(x)^{-1}
                                             \label{U-conds}
\end{eqnarray}
(cf (\ref{U^dagger}) and (\ref{U-extra})) for a given conjugation.
It may be more reasonable to dispense with the last condition in
(\ref{U-conds}). See also the discussion in section IV.A.
\vskip.3cm
\noindent
For the curvature we find
\begin{eqnarray}
F = U^2 = {1 \over 2} \sum_{\epsilon,\mu,\epsilon',\nu} \left[
  U_{\epsilon \mu}(x) \, U_{\epsilon' \nu}(x+\epsilon \, \ell
  \hat{\mu}) - U_{\epsilon' \nu}(x) \, U_{\epsilon \mu}(x+\epsilon' \,
  \ell \hat{\nu}) \right] \, e^{\epsilon \mu} \, e^{\epsilon'\nu} \; .
\end{eqnarray}
The Yang-Mills equation
\begin{eqnarray}
  d^\ast F + A^\natural \bullet F + F \bullet A^\natural =
  U^\natural \bullet (U^2) + (U^2) \bullet U^\natural = 0
\end{eqnarray}
now leads to
\begin{eqnarray}
  U_{\epsilon\mu}(x) &=& {1\over 4n} \, \sum_{\epsilon',\nu} \lbrack
  U_{\epsilon' \nu}(x) \, U_{\epsilon \mu}(x+\ell\epsilon' \hat{\nu})
  \, U_{\epsilon' \nu}(x+\epsilon \ell \hat{\mu})^\dagger
                                           \nonumber \\
 & & + U_{\epsilon'\nu}(x-\epsilon'\ell\hat{\nu})^\dagger \,
  U_{\epsilon\mu}(x-\epsilon'\ell\hat{\nu}) \,
  U_{\epsilon'\nu}(x+\ell(\epsilon\hat{\mu}-\epsilon'\hat{\nu})
  \rbrack
\end{eqnarray}
and the Yang-Mills action takes the form
\begin{eqnarray}
S_{YM} & = & {1 \over 4 \ell^4} \, \mbox{tr} \sum_{a,\epsilon,
 \epsilon',\mu,\nu}
 \left[ 1 - U_{\epsilon\mu}(\ell a) \, U_{\epsilon'\nu}(\ell(a+\epsilon
 \hat{\nu})) \, U_{\epsilon\mu}(\ell(a+\epsilon'\hat{\nu}))^\dagger \,
 U_{\epsilon'\nu}(\ell a)^\dagger \right] \nonumber \\
& = & {1 \over \ell^4} \, \mbox{tr} \sum_{a,\mu,\nu}
 \left[ 1 - U_{+\mu}(\ell a)\, U_{+\nu}(\ell(a+\hat{\mu})) \,
 U_{+\mu}(\ell(a+\hat{\nu}))^\dagger \, U_{+\nu}(\ell a)^\dagger \right]
 \; .
\end{eqnarray}
This is again the Wilson action (cf (\ref{Wilson})). Note, however,
that this result was obtained by imposing an additional constraint,
the second equation in (\ref{U-conds}).

\section{Conclusions}
\setcounter{equation}{0}
We have explored differential algebras on a discrete set $\cal M$. In
section II we introduced 1-forms $e_{ij}, \, i,j \in {\cal M},$ which
generate the differential algebra over $\Cx$. They turned out to be
particularly convenient to work with and, in particular, provided
us with a simple way to `reduce' the
universal differential algebra to smaller differential algebras.
Such `reductions' of the universal differential algebra are
described by certain graphs which can be related to `Hasse diagrams'
determining a locally finite topology. In this way, contact was made
in section III with the recent work by Balachandran et al.
\cite{BBET93} where a calculus on `posets' (partially ordered sets)
has been developed with the idea to discretize continuum models
in such a way that important topological features of continuum physics
(like winding numbers) are preserved. What we learned is that the
adequate framework for doing this is (noncommutative) differential
calculus on discrete sets. As a special example, the differential
calculus which corresponds to an oriented hypercubic lattice graph
reproduces the familiar formalism of lattice (gauge) theories
(see also \cite{DMHS93}). This is, however, just one choice among
many.
\vskip.3cm

In particular, we have studied the differential calculus
associated with the `symmetric' hypercubic lattice graph.
In a certain continuum limit, this calculus tends
to a deformation of the ordinary calculus of differential forms on a
manifold which is known to be related to quantum theory
\cite{DMH92-grav}, stochastics \cite{DMH93-stoch} and, more
exotically, differential calculus on quantum groups \cite{MH+Reut93}.
In the same limit, however, the Yang-Mills action on the symmetric
lattice just tends to the ordinary Yang-Mills action.
\vskip.3cm

In this work, we have presented a formulation of gauge theory on a
discrete set or, more precisely, on graphs describing differential
algebras on a discrete set. This should be viewed as a generalization of
the familiar Wilson loop formulation of lattice gauge theory.
A corresponding gauge theoretical approach to a discrete gravity theory
will be discussed elsewhere \cite{DMH94}. In that case, `symmetric
graphs' play a distinguished role.
\vskip.3cm

As already mentioned in the introduction, it seems that the relation
between differential calculus on finite sets and approximations
of topological spaces established in the present work allows us
to understand the `discrete' gauge theory models of Connes and Lott
\cite{Conn+Lott90} (and many similar models which have been proposed
after their work) as approximations of higher-dimensional gauge
theory models (see \cite{Mant79,Kape+Zoup92}, in particular).
The details have still to be worked out, however.
\vskip.3cm

A differential algebra provides us with a notion of locality
since its graph determines a neighbourhood structure.
If we supply, for example, the set $\Ir$ with the differential algebra
such that $e_{ij} \neq 0$ iff $j=i+1$, then some fixed $i$ and $i+1000$,
say, are quite remote from one another in the sense that they are
connected via many intermediate points.
If, however, we allow also $e_{i (i+1000)} \neq 0$, then the two points
become neighbours. This modification of the graph has crucial
consequences since a nonvanishing $e_{ij}$ yields the possibility of
correlations between fields at the points $i$ and $j$. If a set is
supplied with the universal differential algebra, then
correlations between any two points are allowed and will naturally be
present in a field theory built on it. One could imagine that in such a
field theory certain correlations are dynamically suppressed so that,
e.g., a four-dimensional structure of the set is observed.
\vskip.3cm

The universal differential algebra on a set $\cal M$ corresponds to
the graph where the elements of $\cal M$ are represented
by the vertices and any two points are connected by two (antiparallel)
arrows. To the arrow from $i$ to $j$ we may assign a probability
$p_{ij} \in \lbrack 0 , 1 \rbrack$. We are then dealing with a
`fuzzy graph'. If $p_{ij} \in \lbrace 0,1
\rbrace$ for all $i,j \in {\cal M}$, we recover our concept of a
reduction of the universal differential algebra. A more general
formalism (allowing also values of $p_{ij} \in (0,1)$) could
describe, e.g., fluctuations in the space (-time) dimension
since the latter depends on how many (direct) neighbours a given
site has.
This suggests a quantization of the universal differential
algebra by introducing creation and annihilation operators for the
1-forms $e_{ij}$ (and perhaps higher forms).
\vskip.3cm

After completion of this work we received a preprint by
Balachandran et al \cite{BBELLST94} which also relates (poset)
approximations of topological spaces and noncommutative geometry
although in a way quite different from ours. In particular, they
associate a {\em non}-commutative algebra (of operators on a Hilbert
space) with a poset.

\appendix
\renewcommand{\theequation} {\Alph{section}.\arabic{equation}}

\section*{Relation with \v{C}ech-cohomology}
{}From the universal differential algebra on a discrete set $\cal M$
one can construct a $\Cx$-vector space $\tilde{A}^k({\cal A})$
of {\em antisymmetric} $k$-forms ($k > 0$) generated by
\begin{eqnarray}
   a_{i_0 \ldots i_k} := e_{\lbrack i_0 \ldots i_k \rbrack}
\end{eqnarray}
where $e_{i_0 \ldots i_k}$ is defined in (\ref{e...}) and the square
brackets indicate antisymmetrization.
$\tilde{A}^k({\cal A})$ is {\em not} an $\cal A$-module since
multiplication with $e_i$ leaves the space. For example,
\begin{eqnarray}
  e_i \, (\underbrace{e_{ij} - e_{ji}}_{\in \, \tilde{A}^1({\cal A})})
  = e_{ij} \; \notin \, \tilde{A}^1({\cal A})  \; .
\end{eqnarray}
We set $\tilde{A}({\cal A}) := \bigoplus_{k \geq 0} \tilde{A}^k({\cal
A})$ with $\tilde{A}^0({\cal A}) := {\cal A}$. More generally, one may
consider any reduction $A({\cal A})$ of $\tilde{A}({\cal A})$ obtained
by setting some of the generators $a_{i_0 \ldots i_k}$ to zero. Since
\begin{eqnarray}
   a_{i_0 \ldots i_k} \neq 0 \; \Rightarrow \;
   a_{j_0 \ldots j_\ell} \neq 0  \; \mbox{ if }
   \lbrace j_0, \ldots, j_\ell \rbrace \subset
   \lbrace i_0, \ldots, i_k \rbrace
\end{eqnarray}
one finds from (\ref{dpsi}) that $A({\cal A})$ is closed under $d$.
Now
\begin{eqnarray}
   a_{i_0 \ldots i_k} = e_{\lbrack i_0} \otimes \cdots \otimes
   e_{i_k \rbrack}
\end{eqnarray}
(cf the second remark in section II) yields a representation of
$A({\cal A})$. If $f \in A^k({\cal A})$, then
$f = \sum_{j_0 \ldots j_k} a_{j_0 \ldots j_k} \, f_{j_0 \ldots j_k}$
with antisymmetric coefficients $f_{j_0 \ldots j_k} \in \Cx$ and thus
\begin{eqnarray}
   f(i_0, \ldots, i_k) = f_{i_0 \ldots i_k}  \; .
\end{eqnarray}
Furthermore, from (\ref{dpsi}) we obtain
\begin{eqnarray}
    d f(i_0, \ldots, i_{k+1}) = \sum_{j=0}^{k+1} (-1)^j \,
        f(i_0, \ldots, i_{j-1}, \widehat{i_j}, i_{j+1}, \ldots i_k)
                                \label{d-Cech}
\end{eqnarray}
where a hat indicates an omission. These formulae are reminiscent
of \v{C}ech cohomology theory. The relation will be explained in the
following.
\vskip.3cm

Let ${\cal U} = \lbrace {\cal U}_i \, \mid \, i \in {\cal M} \rbrace$
be an open covering of a manifold $M$.
In \v{C}ech cohomology theory a $k$-{\em simplex} is any ($k+1$)-tuple
$(i_0,i_1, \ldots, i_k)$ such that ${\cal U}_{i_0} \cap \ldots \cap
{\cal U}_{i_k} \neq \emptyset$. A \v{C}ech-{\em cochain} is any
(totally) antisymmetric mapping
\begin{eqnarray}
    f \, : \, (i_0, \ldots , i_k) \mapsto f(i_0, \ldots , i_k)
                          \in {\cal K}
\end{eqnarray}
where ${\cal K}= \Cx, \Rl, \Ir$. The set of all $k$-cochains forms
a $\cal K$-linear space $C^k({\cal U}, {\cal K})$. The
\v{C}ech-coboundary operator $d \, : \, C^k({\cal U}, {\cal K})
\rightarrow C^{k+1}({\cal U}, {\cal K})$ is then defined by
(\ref{d-Cech}).
If ${\cal U}$ is a {\em good} covering, then the de Rham cohomology of
the manifold is isomorphic to the \v{C}ech cohomology with ${\cal K}
= \Rl$ \cite{Bott+Tu82}. A covering of a manifold is `good' if all
finite nonempty intersections are contractible.
\vskip.3cm

This suggests the following way to associate a topology with
$A({\cal A})$. For each element $i \in \cal M$ we have an open
set ${\cal U}_i$. Intersection relations are then determined by
\begin{eqnarray}
   a_{i_0 \ldots i_k} \neq 0 \; \Leftrightarrow \;
   {\cal U}_{i_0} \cap \ldots \cap {\cal U}_{i_k} \neq \emptyset
     \; .
\end{eqnarray}
\vskip.3cm

In algebraic topology one
constructs from the intersection relations of the open sets ${\cal U}_i$
a simplicial complex, the {\em nerve} of $\cal U$ (see
\cite{Bott+Tu82}, for example). If ${\cal U}_i \cap {\cal U}_j \neq
\emptyset$, we connect the vertices $i$ and $j$ with an edge. Since the
intersection relation is symmetric we may also think of drawing two
antiparallel arrows between $i$ and $j$ (thus making contact with
the procedure in section III).
A triple intersection relation ${\cal U}_i \cap {\cal U}_j \cap
{\cal U}_k$ (where $i,j,k$ are pairwise different) corresponds to the
face of the triangle with corners $i,j,k$, and so forth.
Instead of the simplicial complex (the nerve) obtained in this way
-- which need not be a simplicial approximation of the manifold
(see \cite{Hock+Youn61}) -- we can construct a
Hasse diagram with the same information as follows. The first row
consists of the basic vertices corresponding to the elements of $\cal M$
respectively the open sets ${\cal U}_i$, $i \in {\cal M}$. The next row
(below the first) consists of vertices associated with the nontrivial
intersections of pairs of the open sets ${\cal U}_i$. The vertex
representing ${\cal U}_i \cap {\cal U}_j \neq \emptyset$ then gets
connections with ${\cal U}_i$ and ${\cal U}_j$. With each
${\cal U}_i \cap \cdots \cap {\cal U}_j \neq \emptyset$ we associate a
new vertex and proceed in an obvious way.
\vskip.3cm

In section III we started from a differential calculus (a reduction
of the universal differential calculus on $\cal M$) and derived a
Hasse diagram from it which then determined a covering of some
topological space. The covering defines a \v{C}ech complex and we
have seen above that the latter is represented by some space
$A({\cal A})$ of antisymmetric forms.


\small

\begin{references}
\bibitem{Conn86} A. Connes, ``Non-commutative differential
 geometry", Publ. I.H.E.S. {\bf 62}, 257 (1986).
\bibitem{Coqu89} R. Coquereaux, ``Noncommutative geometry and
 theoretical physics", J. Geom. Phys. {\bf 6}, 425 (1989).
\bibitem{MH+Reut93} F. M\"uller-Hoissen and C. Reuten,
 ``Bicovariant differential calculus on $GL_{p,q}(2)$ and quantum
 subgroups",  J. Phys. A (Math. Gen.) {\bf 26}, 2955 (1993);
 F. M\"uller-Hoissen, ``Differential calculi on quantum (sub-)
 groups and their classical limit", preprint, G\"ottingen University,
 hep-th/9401152, to appear in the proceedings of the workshop on
 {\it Generalized Symmetries in Physics}, ASI Clausthal, July 1993.
\bibitem{Conn+Lott90} A. Connes and J. Lott, ``Particle models and
 noncommutative geometry", Nucl. Phys. B (Proc. Suppl.) {\bf 18},
 29 (1990);  ``The metric aspect of noncommutative geometry", in
 {\it New Symmetry Principles in Quantum Field Theory}, edited by
 J. Fr\"ohlich et al. (Plenum Press, New York, 1992) p. 53.
\bibitem{DMH92} A. Dimakis and F. M\"uller-Hoissen,
 ``Quantum mechanics on a lattice and $q$-deformations",
 Phys. Lett. {\bf 295B}, 242 (1992).
\bibitem{DMHS93} A. Dimakis, F. M\"uller-Hoissen and T. Striker,
 ``From continuum to lattice theory via deformation of the differential
 calculus", Phys. Lett. {\bf 300B}, 141 (1993);
 ``Noncommutative differential calculus and lattice gauge theory",
 J. Phys. A {\bf 26}, 1927 (1993).
\bibitem{DMH93-fs} A. Dimakis and F. M\"uller-Hoissen,
 ``Differential calculus and gauge theory on finite sets",
 J. Phys. A: Math. Gen. {\bf 27}, 3159 (1994);
 ``Differential calculus and discrete structures", preprint,
 G\"ottingen University, hep-th/9401150, to appear in the proceedings
 of the workshop on  {\it Generalized Symmetries in Physics}, ASI
 Clausthal, July 1993.
\bibitem{discrete-spacetime}
 V. Ambarzumian and D. Iwanenko, ``Zur Frage nach Vermeidung der
 unendlichen Selbstr\"uckwirkung des Elektrons", Z. Physik
 {\bf 64}, 563 (1930);
 A. Ruark, ``The roles of discrete and continuous theories in
 physics", Phys. Rev. {\bf 37}, 315 (1931);
 H.S. Snyder, ``Quantized space-time", Phys. Rev. {\bf 71}, 38 (1947);
 D. Finkelstein, ``Space-time code", Phys. Rev. {\bf 184}, 1261 (1969);
 R.P. Feynman, ``Simulating physics with computers",
 Int. J. Theor. Phys. {\bf 21}, 467 (1982);
 M. Minsky, ``Cellular vacuum", Int. J. Theor. Phys. {\bf 21}, 537
 (1982);
 G. Feinberg, R. Friedberg, T.D. Lee and H.C. Ren,
 ``Lattice gravity near the continuum",
 Nucl. Phys. B {\bf 245}, 343 (1984);
 T.D. Lee, ``Discrete mechanics", in {\it How far are we from the
 gauge forces}, edited by A. Zichichi (Plenum Press, New York, 1985)
 p.15;
 H. Yamamoto, ``Quantum field theory on discrete
 space-time", Phys. Rev. {\bf 30}, 1727 (1984);
 ``Quantum field theory on discrete space-time. II", Phys. Rev.
 {\bf 32}, 2659 (1985);
 H. Yamamoto, A. Hayashi, T. Hashimoto and M. Horibe,
 ``Towards a canonical formalism of field theory on discrete
 spacetime", preprint, Fukui University, Japan (1993);
 L. Bombelli, J. Lee, D. Meyer and R.D. Sorkin,
 ``Space-time as a causal set", Phys. Rev. Lett. {\bf 59}, 521 (1987);
 G. 't Hooft, ``Quantization of discrete deterministic
 theories by Hilbert space extension", Nucl. Phys. B {\bf 342}, 471
 (1990);
 D. Finkelstein and J.M. Gibbs, ``Quantum relativity",
 Int. J. Theor. Phys. {\bf 32}, 1801 (1993).
\bibitem{Sork91} R.D. Sorkin, ``Finitary substitute for
 continuous topology", Int. J. Theor. Phys. {\bf 30}, 923 (1991).
\bibitem{Anez94} C. Aneziris, ``Topology and statistics in zero
 dimensions'', Int. J. Theor. Phys. {\bf 33}, 535 (1994).
\bibitem{Kape+Zoup92}
 D. Kapetanakis and G. Zoupanos, ``Coset space dimensional
 reduction of gauge theories", Phys. Rep. {\bf 219}, 1 (1992).
\bibitem{Mant79} N.S. Manton, ``A new six-dimensional approach
 to the Weinberg-Salam model", Nucl. Phys. B {\bf 158}, 141 (1979).
\bibitem{Conn93} A. Connes, ``Noncommutative geometry and
 physics", Les Houches lecture notes, IHES/M/93/32.
\bibitem{Ritz08} W. Ritz, ``\"Uber ein neues Gesetz der
 Serienspektren", Phys. Z. {\bf 9}, 521 (1908);
 J.R. Rydberg, ``La distribution des raies spectrales", in
 {\it Rapports pr{\'e}sent{\'e}s au Congr{\`e}s International de
 Physique}, Vol. 2 (Gauthier-Villars, Paris, 1900), p. 200.
\bibitem{Born+Jord25} M. Born and P. Jordan, ``Zur
 Quantentheorie", Zeitschr. f. Physik {\bf 34}, 858 (1925).
\bibitem{Hock+Youn61} J.G. Hocking and G.S. Young, {\em Topology}
 (Addison-Wesley, Reading MA, 1961).
\bibitem{Evak94} A.V. Evako, ``Dimension on discrete spaces",
 preprint, Syracuse University, gr-qc/9402035.
\bibitem{BBET93} A. P. Balachandran, G. Bimonte, E. Ercolessi and
 P. Teotonio-Sobrinho, ``Finite approximations to quantum physics:
 quantum points and their bundles", preprint, Syracuse University,
 SU-4240-550 (1993).
\bibitem{Kast88} D. Kastler, {\em Cyclic cohomology within the
 differential envelope} (Hermann, Paris, 1988).
\bibitem{Rabi81} J. Rabin, ``Homology theory of lattice fermion
 doubling", Nucl.Phys. {\bf B201}, 315 (1981);
 A. Guth, ``Existence proof of a nonconfining phase
 in four-dimensional $U(1)$ lattice gauge theory", Phys. Rev D {\bf
 21}, 2291 (1980).
\bibitem{DMH92-grav} A. Dimakis and F. M\"uller-Hoissen,
 ``A noncommutative differential calculus and its relation to
 gauge theory and gravitation",
 Int. J. Mod. Phys. A (Proc. Suppl.) {\bf 3A}, 474 (1993);
 ``Noncommutative differential calculus, gauge theory and
 gravitation", preprint, G\"ottingen University, GOE--TP 33/92 (1992).
\bibitem{DMH93-stoch} A. Dimakis and F. M\"uller-Hoissen,
 ``Stochastic differential calculus, the Moyal $\ast$-product, and
 noncommutative geometry", Lett. Math. Phys. {\bf 28}, 123 (1993);
 ``Noncommutative Differential Calculus: Quantum Groups, Stochastic
 Processes, and the Antibracket", preprint, G\"ottingen University,
 GOET-TP 55/93, hep-th/9401151.
\bibitem{DMH94} A. Dimakis and F. M\"uller-Hoissen,
 ``Noncommutative geometry and gravitation", in preparation.
\bibitem{BBELLST94} A. P. Balachandran, G. Bimonte, E. Ercolessi,
 G. Landi, F. Lizzi, G. Sparano and P. Teotonio-Sobrinho, ``Finite
 quantum physics and noncommutative geometry",
 preprint, Syracuse University, SU-4240-567, March 1994.
\bibitem{Bott+Tu82} R. Bott and L.W. Tu, {\em Differential Forms in
 Algebraic Topology} (Springer, New York, 1982).
\end{references}

\normalsize
\end{document}